\pdfoutput=1
\documentclass{article}
\usepackage[numbers]{natbib}
\usepackage[preprint]{neurips_2022}
\usepackage{bm}
\usepackage[utf8]{inputenc} 
\usepackage[T1]{fontenc}    
\usepackage{hyperref}       
\usepackage{url}
\usepackage{booktabs}       
\usepackage{amsfonts}       
\usepackage{nicefrac}       
\usepackage{microtype}      
\usepackage{xcolor}         
\usepackage{subcaption}
\usepackage{selectp}  
\title{\mytitle}

\author{
Yongtao Wu
    \\
 KTH Royal Institute of Technology\\
  \texttt{yongtaowu98@gmail.com} \\
   \And
   Grigorios G Chrysos \\
   EPFL, Switzerland \\
   \texttt{grigorios.chrysos@epfl.ch} \\
   \AND
   Volkan Cevher \\
EPFL, Switzerland \\
   \texttt{volkan.cevher@epfl.ch} \\
}

\usepackage{multirow}
\usepackage[normalem]{ulem}
\usepackage{amsmath}
\usepackage{amssymb}
\usepackage{mathtools}
\usepackage{amsthm}
\usepackage[capitalize,noabbrev]{cleveref}
\theoremstyle{plain}
\newtheorem{theorem}{Theorem}[section]

\newtheorem{lemma}[theorem]{Lemma}

\theoremstyle{definition}

\theoremstyle{remark}

\usepackage[textsize=tiny]{todonotes}

\graphicspath{ {figures/} }
\newcommand{\minus}{\scalebox{0.75}[1.0]{$-$}}
\providecommand{\realnum}{\mathbb{R}}
\providecommand{\complexnum}{\mathbb{C}}
\providecommand{\naturalnum}{\mathbb{N}}
\providecommand{\bmcal}[1]{\bm{\mathcal{#1}}}

\providecommand{\matnot}[1]{_{[{#1}]}}  
\providecommand{\invar}{x}  
\providecommand{\prodbias}{h}  
\providecommand{\binvar}{\boldsymbol{\widetilde{\invar}}}  
\providecommand{\outvar}{y}  
\providecommand{\boutvar}{\boldsymbol{
\widetilde{\outvar}}}

\providecommand{\compoly}{APOLLO}
\providecommand{\pinet}{$\Pi$-Nets}
\providecommand{\resnet}{ResNet}

\providecommand{\modelcccp}{$\mathbb{C}$FC}

\providecommand{\biasmodelrccp}{$\mathbb{R}$BC}
\providecommand{\biasmodelrncp}{$\mathbb{R}$BN}

\providecommand{\biasmodelcccp}{$\mathbb{C}$FBC}
\providecommand{\biasmodelcncp}{$\mathbb{C}$FBN}

\providecommand{\biasmodelmixccp}{$\mathbb{C}$BC}
\providecommand{\biasmodelmixncp}{$\mathbb{C}$BN}

\providecommand{\sgamma}[1]{r_{\scalebox{.4}{#1}}}

\newcommand{\mytitle}
{Adversarial Audio Synthesis with Complex-valued Polynomial Networks}
\usepackage{pifont}
\newcommand{\xmark}{\textcolor{red}{\ding{55}}}
\definecolor{mygreen}{RGB}{30, 180, 50}
\newcommand{\colorcheck}{\textcolor{mygreen}{\pmb{\checkmark}}}

\begin{document}
\maketitle

\begin{abstract}
\label{sec:complex_poly_abstract}

Time-frequency (TF) representations in audio synthesis have been increasingly modeled with real-valued networks. However, overlooking the complex-valued nature of TF representations can result in suboptimal performance and
require additional modules (e.g., for modeling the phase). To this end, we introduce complex-valued polynomial networks, called \compoly{}, that integrate such complex-valued representations in a natural way. Concretely, \compoly{} captures high-order correlations of the input elements using high-order tensors as scaling parameters. By leveraging standard tensor decompositions, we derive different architectures and enable modeling richer correlations. We outline such architectures and showcase their performance in audio generation across four benchmarks. As a highlight, \compoly{} results in $17.5\%$ improvement over adversarial methods and $8.2\%$ over the state-of-the-art diffusion models on SC09 dataset in audio generation. Our models can encourage the systematic design of other efficient architectures on the complex field.
\end{abstract} \section{Introduction}
\label{sec:complex_poly_introduction}

Generative Adversarial Networks (GANs) enable photo-realistic synthesis in image-related tasks~\citep{NIPS2014_5ca3e9b1,pix2pix2017,karras2018style, brock2018large,chrysos2021conditional}. Their stellar performance has prompted their use in unconditional audio synthesis, which aims to synthesize consistent utterances from noise~\citep{donahue2018adversarial,engel2019gansynth,marafioti2019adversarial,palkama20_interspeech}. However, the human perception is sensitive to both global and local coherence of the waveform~\citep{engel2019gansynth}, which makes audio synthesis an inherently challenging task. We argue that the design choices, i.e., the audio representations and the network architecture, hold a key role in successful audio synthesis.

Raw waveform is primarily used for unconditional speech generation~\cite{donahue2018adversarial}. Recent studies focus on the two-dimensional time-frequency (TF) representation due to its both increased performance and the theoretical expressivity~\cite{5952091,jordi_pons_2018_1492497,marafioti2019adversarial}. In the TF representation, the raw waveform is transformed through the Short-time Fourier transform (STFT) or Constant-Q transform (CQT) to the frequency domain, which is naturally expressed with complex numbers. To avoid using complex numbers often the phase information is discarded and only the magnitude is maintained. However, dropping the phase deteriorates the performance and results in a lack of phase coherence in synthesized audio~\cite{engel2019gansynth}. Importantly, without the phase information, the TF representation is not invertible. This raises the question: \emph{How can we explicitly model the complex-valued TF representation?}

A complex number can be expressed using two real numbers, so a real-valued neural network (RVNN), i.e., a typical feed-forward neural network, with two real-valued outputs could represent complex numbers.
The TF representation can then be thought of as a two-channel image, which enables us to utilize the progress made on RVNNs in the previous decade~\cite{7780459}. 
However, a more natural representation is to design complex-valued neural networks (CVNNs) with complex-valued weights or complex-valued inputs/outputs. Incidentally, CVNNs have demonstrated higher generalization ability~\citep{hirose2011comparison}, which is partly attributed to the degree of freedom at the synaptic weighting of two-channel RVNNs being increased due to complex multiplication~\cite{bassey2021survey}. Even though the richer capacity has led to an increasing attention on CVNNs~\cite{trabelsi2018deep,choi2018phase,yang2020complex}, CVNNs have yet to demonstrate state-of-the-art performance on audio generation.

\begin{figure*}[t]
    \centering
    \includegraphics[width=1\linewidth]{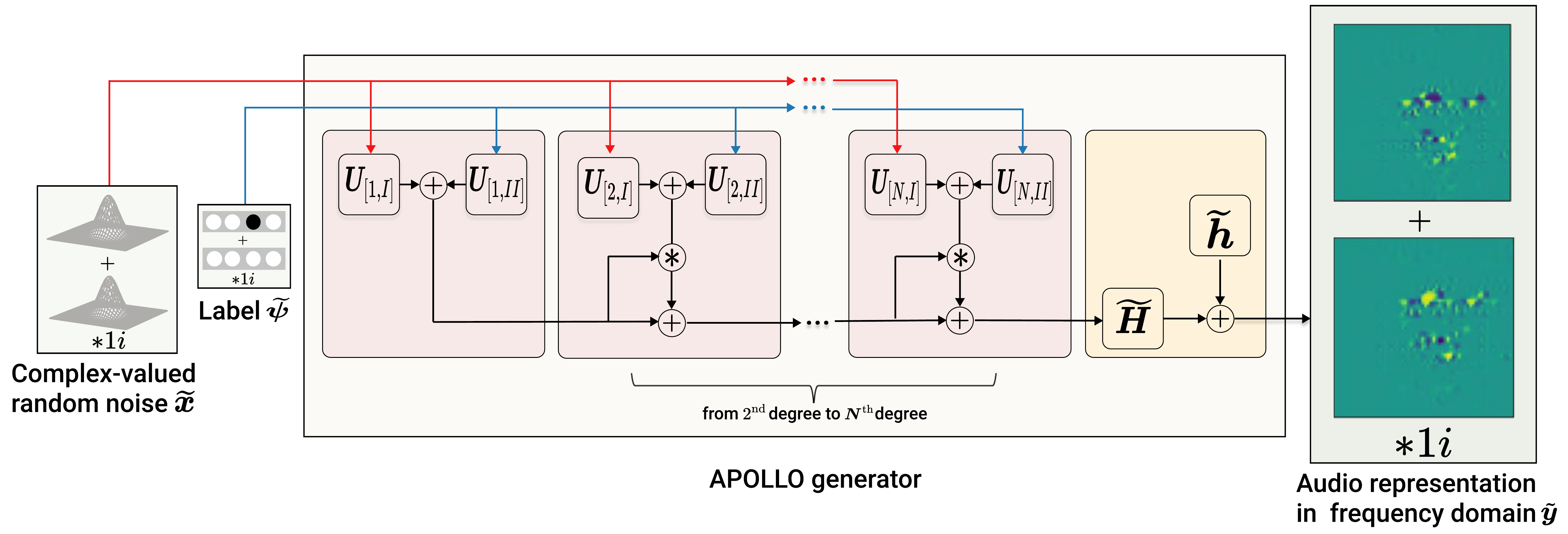}
    \vspace{-4mm}
\caption{
In this work we propose a class of functions, called \compoly{}, where the complex-valued output is a polynomial of the complex-valued input. The input of the \compoly{} generator is complex-valued noise and the output is the representation of audio in the frequency domain (e.g., STFT, CQT). \compoly{} can also receive conditional information, e.g., class-label, which can be treated as a complex-valued one-hot vector with zero imaginary part. The generator in the schematic corresponds to \eqref{eq:complex_poly_model1_rec}.  All learnable parameters inside pink blocks (yellow blocks) are real-valued (complex-valued).
}
\label{fig:prodpoly_model_conditional}
\end{figure*}

On the other hand, recent theoretical advances~\citep{
Jayakumar2020Multiplicative, fan2021expressivity} prove that models with second-degree polynomials enlarge the set of functions that can be represented exactly with zero error. Polynomial nets have demonstrated flexibility and efficiency over standard neural networks in various tasks, e.g., image generation~\citep{karras2018style,chrysos2021conditional}, image recognition~\citep{Wang_2018_CVPR}, reinforcement learning~\citep{Jayakumar2020Multiplicative}, and sequence modelings~\citep{su2020convolutional}. This motivates us to design a class of polynomial nets, called \compoly, that extracts complex-valued representations for audio generation. We utilize complex-valued polynomial networks that express a complex-valued output as a high-degree polynomial expansion of the complex-valued input, as illustrated in Fig.~\ref{fig:prodpoly_model_conditional}. The unknown parameters of the expansion are naturally represented as tensors, and we use standard tensor decompositions for reducing the number of learnable parameters. We determine how specific decompositions result in simple recursive formulations that enable us to implement arbitrary degree polynomial expansions efficiently. 
We conduct a number of experiments that exhibit the efficacy of \compoly{} in unconditional (conditional) audio generation and multimodal generation. Overall, our contributions can be summarized as follows:
\begin{itemize}
    \vspace{-0.5em}
    \setlength\itemsep{-0.2em}
    \item We introduce a new class of complex-valued polynomial neural networks. We reveal how different architectures can be obtained by changing the factorization of the unknown parameters in the polynomial expansion. 
    \item We conduct a thorough evaluation on 
    audio generation and showcase the advantage of \compoly{} when compared with the prior art.
    Through directly modelling the complex-valued TF representation, \compoly{} is free of phase estimation. 
    \item \compoly{} is extended in case of multiple inputs, e.g., when additional variables are available. We implement such a polynomial expansion on conditional audio generation. Additionally, we investigate the efficacy of learning shared representations on multimodal generation.
    \item  Lastly, our model can also be applied in additional audio-related tasks, e.g., speech recognition and speech enhancement.
\end{itemize}
\vspace{-2pt}

We believe that our results can further encourage the research community to consider the complex representations, especially in the context of polynomial networks. Therefore, we will make the source code available upon the acceptance of the paper to accelerate such experimentation. 

\section{Related work}
\label{sec:complex_poly_related}
\textbf{Audio generation.} 
Unconditional audio generation aims to generate audio creatively, which has practical significance in the real world
~\citep{marafioti2019adversarial,donahue2018adversarial,engel2019gansynth,palkama20_interspeech}. WaveGAN and SpecGAN~\cite{donahue2018adversarial} are the first attempts to unconditionally synthesize audio using GANs. WaveGAN directly models the raw audio while SpecGAN models the log spectrum of the STFT in the frequency domain. 
The drawback of SpecGAN includes the demand for phase recovery with Griffin-Lim ~\cite{griffin1984signal} during synthesis step. GANSynth~\cite{engel2019gansynth} improves the previous work by jointly modeling the mel-spectrogram and the instantaneous frequency, which results in more consistent harmonics and avoids additional phase estimation technique.
However, the phase misalignment in GANSynth leads to energy loss due to the cancellation of frequency bands~\cite{marafioti2019adversarial}. Similar to SpecGAN, TiFGAN~\cite{marafioti2019adversarial} only models the log spectrum of the STFT while it estimates the phase by Phase-gradient heap integration (PGHI)~\cite{pruvsa2016real}, which avoids expensive iterative Griffin-Lim in SpecGAN and magnitude integration in GANSynth.
\citet{Gunasekaran2020ImprovedSS} improve the conditional spoken digit generation using mel-spectrogram and Griffin-Lim.
\citet{haque2020guided} propose GGAN to achieve high fidelity audio generation with a fewer labelled audio clips, which rely on log spectrum representation and PGHI recovery. 
Contrary to most aforementioned methods, our \compoly{} directly models the STFT in the complex domain without using phase recovery.
Apart from adversarial generation, diffusion-based models (DiffWave~\citep{kong2021diffwave}) and state space~\citep{gu2022efficiently} based models (SASHIMI~\citep{goel2022s}) are also receiving increasing attention. Compared with adversarial methods, DiffWave results in slower inference speed due to the reverse process. 
Lastly, text-to-speech (TTS) systems, as another category of audio generation, generate the audio representation based on the text information and then utilize a neural vocoder to obtain the waveform. Many solutions are proposed in recent years~\citep{vandenoord16_ssw,pmlrv80kalchbrenner18a,wang2017tacotron,shen2018natural,valle2021flowtron}. Both SASHIMI and DiffWave can also be implemented as neural vocoder and achieve excellent performance against WaveNet~\citep{vandenoord16_ssw}.

\textbf{Complex-valued neural networks (CVNNs).}
The ideas behind CVNNs can be traced back to at least \citet{5726778} that derive backpropagation rules with respect to a real and an imaginary part.
The mathematical motivation as well as the biological feasibility of CVNNs has been explored during the previous decade~\citep{reichert2013neuronal,tygert2016mathematical}.
Being equipped with a rich representational capacity in the complex field, CVNNs have drawn increasing attention in a wide range of applications. \citet{oyallon2015deep} design a deep scattering convolution network to improve the performance of image descriptors. 
\citet{trouillon2016complex} use complex-valued embeddings for simple link prediction. In audio-related tasks, the noise speech is enhanced by being converted and passed through STFT, complex-valued network, and inverse STFT~\cite{choi2018phase}.
Another line of research applies complex operations on top of typical neural networks. \citet{arjovsky2016unitary} use complex operations on recurrent neural network to circumvent problems of vanishing and exploding gradients.
\citet{trabelsi2018deep} propose complex convolutional networks and algorithms for complex batch-normalization, complex weight initialization, and complex activations.

\textbf{Polynomial neural networks (PNNs).} 
PNNs have been studied for several decades ~\cite{ivakhnenko1971polynomial,giles1987learning,shin1991pi,oh2003polynomial,li2003sigma,xiong2007training}. However, many of the aforementioned PNNs do not scale well for high-dimensional data samples, such as those required for modern generative models. 
\citet{chrysos2019polygan} introduce a high-degree polynomial generator, which approximates the distribution of high-dimensional data. The work was subsequently extended on a range of applications in the class of function approximators called \pinet, where \citet{9353253} exhibit the increased expressivity of real-valued PNNs. 
Lastly, \citet{DBLP:journals/corr/abs-1905-05605} focus on approximating the operations such as sigmoid, ReLU, and max pooling with polynomial operations.
Differently from previous works, we explicitly propose polynomial expansions on the complex field. \section{Methods}
\label{sec:complex_poly_methods}

\textbf{Notation}
We introduce the core operators/symbols, while a more detailed notation is deferred to Sec.\ref{sec:complex_poly_detailed_notation} in the appendix. Real-valued vectors/matrices/tensors are symbolized by lowercase/uppercase/calligraphic boldface letters, e.g., $\boldsymbol{y}, \boldsymbol{Y}, \boldsymbol{\mathcal{Y}}$. 
All complex-valued variables are symbolized with wide tilde, e.g., $\boldsymbol{\widetilde{y}}$, $\boldsymbol{\widetilde{Y}}$, $\widetilde{\boldsymbol{\mathcal{Y}}}$.
The mode-$m$ vector product between a complex-valued tensor and multiple complex-valued vectors is denoted as: $\widetilde{\boldsymbol{\mathcal{Y}}} \times_{1} \widetilde{\boldsymbol{u}}^{(1)} \times_{2} \widetilde{\boldsymbol{u}}^{(2)} \times_{3} \cdots \times_{M} \widetilde{\boldsymbol{u}}^{(M)} =\widetilde{\boldsymbol{\mathcal{Y}}} \prod_{m=1}^{M} \times_{m} \widetilde{\boldsymbol{u}}^{(m)}.$
Khatri-Rao product and Hadamard product are denoted by $\odot$ and $ *$, respectively. 

\subsection{Complex Polynomial Neural Networks}
According to Mergelyan's
Theorem~\cite{rudin1987real}, any smooth complex-valued function could be approximated by a polynomial. A complex polynomial refers to a polynomial expansion with complex-valued coefficients.
Our goal is to learn an $N^{\text{th}}$ degree polynomial expansion with respect to the input $\widetilde{\boldsymbol{x}}\in \mathbb{C}^{d}$ with an $o-$dimensional output $\boldsymbol{\widetilde{y}}$ as follows:
\begin{equation}
\boldsymbol{\widetilde{y}}
=\sum_{n=1}^{N}\left(\boldsymbol{\mathcal{\widetilde{W}}}^{[n]} \prod_{j=2}^{n+1} \times{ }_{j} \widetilde{\boldsymbol{x}}\right)+\boldsymbol{\widetilde{\prodbias}},
\label{equ:3_3}
\end{equation}
    where  $\left\{\boldsymbol{\mathcal{\widetilde{W}}}^{[n]} \in \mathbb{C}^{o \times\prod_{m=1}^{n}{\times}_{m} d} \right\}_{n=1}^{N}$ and $\boldsymbol{\widetilde{\prodbias}} \in \mathbb{C}^{o}$ are learnable parameters.
    The drawback of \eqref{equ:3_3} is that it requires an exponential number of parameters for high-degree polynomial expansions, which is not scalable. Instead, we will apply coupled CP decomposition with factor sharing to reduce the learnable parameters~\cite{kolda2009tensor,9353253}.
\label{sec:model_comlex_valued_coefficients}
Firstly, we introduce polynomial expansions with complex-valued input variables and complex-valued coefficients, where the real and the imaginary coefficients are modelled with independent variables. Sequentially, we make additional assumptions that can further reduce the parameters and lead to efficient implementations. We illustrate below the derivation of \modelcccp{} for third-degree expansion, while we stress out that the recursive formulation can be applied for \emph{any arbitrary degree polynomial expansion}.

\textbf{\modelcccp{} (Fully coupled decomposition)} In the first model we assume all the weight tensors $\{\boldsymbol{\mathcal{\widetilde{W}}}^{[n]} \}_{n=1}^{N}$ of \eqref{equ:3_3} are jointly factorized by a coupled CP decomposition to share factors between different degrees. For a third-degree expansion that becomes:
\begin{itemize} 
    \vspace{-1.1em}
    \setlength\itemsep{-0.3em}
    \item First degree parameters: $\boldsymbol{\widetilde{W}}^{[1]} = \boldsymbol{\widetilde{H}}\boldsymbol{\widetilde{U}}\matnot{1}^T$.
    \item Second degree parameters: $\boldsymbol{\widetilde{W}}^{[2]}_{(1)} = \boldsymbol{\widetilde{H}}(\boldsymbol{\widetilde{U}}\matnot{3} \odot \boldsymbol{\widetilde{U}}\matnot{1})^T + \boldsymbol{\widetilde{H}}(\boldsymbol{\widetilde{U}}\matnot{2} \odot \boldsymbol{\widetilde{U}}\matnot{1})^T$.
    
    \item Third degree parameters: $\boldsymbol{\widetilde{W}}^{[3]}_{(1)} = \boldsymbol{\widetilde{H}}(\boldsymbol{\widetilde{U}}\matnot{3} \odot \boldsymbol{\widetilde{U}}\matnot{2} \odot \boldsymbol{\widetilde{U}}\matnot{1})^T $.
\end{itemize}
By leveraging the aforementioned factorization and utilizing 
simple algebra, a simple recursive form can be retained (which is also generalizable to an $N^{\text{th}}$ degree polynomial) as we prove in Sec.\ref{ssec:sup_for_cccp}:
\begin{equation}
\boldsymbol{\widetilde{y}}_{n}=\left(\boldsymbol{\widetilde{U}}_{[n]}^{T} \widetilde{\boldsymbol{x}}\right) * \boldsymbol{\widetilde{y}}_{n-1}+\boldsymbol{\widetilde{y}}_{n-1},
\label{eq:polygan_recursive_gen_third_order_cccp_final}
\end{equation}
for $n=2, \ldots, N$ with $\boldsymbol{\widetilde{y}}_{1}=\boldsymbol{\widetilde{U}}_{[1]}^{T} \widetilde{\boldsymbol{x}}$ and $\boldsymbol{\widetilde{y}}=\boldsymbol{\widetilde{H}} \boldsymbol{\widetilde{y}}_{N}+\boldsymbol{\widetilde{\prodbias}}$, where 
$ \boldsymbol{\widetilde{U}}\matnot{n} \in \complexnum^{d\times k},
\boldsymbol{\widetilde{H}}  \in \complexnum^{o\times k},
\bm{\widetilde{h}} \in 
\complexnum^{o}.
$
Intuitively, each recursive term increases the degree of expansion by one, using three fundamental operations: a) an affine transformation with the input, i.e., $\boldsymbol{\widetilde{U}}_{[n]}^{T} \widetilde{\boldsymbol{x}}$, b) an element-wise product with the previous term, c) a skip connection with the previous term. Those three operations are easy to implement and enable any arbitrary degree of expansion to be achieved. 

\textbf{\biasmodelcccp{}~(Fully coupled decomposition with bias)}
\label{ssec:biasmodelcccp}
In this section, we develop a new decomposition to allow the existence of bias term 
$\bm{\widetilde{\rho}}\matnot{n} \in 
\complexnum^{k}$
compared with the previous model, which can increase the expressivity of model in practice. The recursive relation is as follows (proof in Sec.\ref{ssec:sup_biasmodelcccp}) 
\begin{equation}
    \boutvar_n = (\bm{\widetilde{U}}\matnot{n}^T \binvar + \bm{\widetilde{\rho}}\matnot{n}) * \boutvar_{n-1} \;, 
    \label{eq:biasmodelcccp}
\end{equation}
for $n=2, \ldots, N$ with $\boldsymbol{\widetilde{y}}_{1}=\boldsymbol{\widetilde{U}}_{[1]}^{T} \widetilde{\boldsymbol{x}}$ and $\boldsymbol{\widetilde{y}}=\boldsymbol{\widetilde{H}} \boldsymbol{\widetilde{y}}_{N}+\boldsymbol{\widetilde{\prodbias}}$.

\textbf{\biasmodelmixccp{}~(Coupled decomposition with bias)}
\label{ssec:biasmodelmixccp}
The model above assumes different parameters for the real and the imaginary part, however, we could also perform sharing of certain factors between the real and the imaginary part to reduce the learnable parameters.
Even though multiple sharing of factors could be implemented, we have assumed that every factor is shared apart from the factor corresponding to the unfolding dimension. That is, all terms between the real and the imaginary coefficients are shared apart from the matrix $\boldsymbol{\widetilde{H}}$.
 The following recursive relationship between different degrees is obtained: (proved in Sec.\ref{ssec:sup_for_lccp}):
\begin{equation}
    \boutvar_n = (\bm{U}\matnot{n}^T \binvar + \bm{\rho}\matnot{n}) * \boutvar_{n-1} \;, 
    \label{eq:polygan_recursive_gen_third_order_lccp_final}
\end{equation}
for $n=2, \ldots, N$ with $\boldsymbol{\widetilde{y}}_{1}=\boldsymbol{U}_{[1]}^{T} \widetilde{\boldsymbol{x}}$ and $\boldsymbol{\widetilde{y}}=\boldsymbol{\widetilde{H}} \boldsymbol{\widetilde{y}}_{N}+\boldsymbol{\widetilde{\prodbias}}$.
Compared with the recursive formulation~\eqref{eq:biasmodelcccp} in the previous decomposition, we observe that \eqref{eq:polygan_recursive_gen_third_order_lccp_final} reduces the number of parameters by converting
$ \boldsymbol{\widetilde{U}}\matnot{n} \in \complexnum^{d\times k}$ to 
$ \boldsymbol{U}\matnot{n} \in \realnum^{d\times k}$
and 
$\bm{\widetilde{\rho}}\matnot{n} \in 
\complexnum^{k}$
to 
$\bm{\rho}\matnot{n} \in 
\realnum^{k}$ among each degree.

\subsection{Other model variants} 
\label{sec:model_real_valued_coefficients}
To obtain more expressive models, we can further apply a joint hierarchical decomposition instead of separating the interactions between different
degrees or add a shortcut connection embedded into the relationship, which is motivated by the skip connections in ResNet \cite{7780459}.
In addition, the aforementioned models assume polynomial with complex-valued coefficients.
However, we can also simplify the models and implement them with real-valued coefficients, 
i.e., we could replace $\boldsymbol{\widetilde{H}}$ with real-valued matrix $\boldsymbol{H}$, $\boldsymbol{\widetilde{h}}$ with real-valued vector $\boldsymbol{h}$. Due to the lack of space, we defer these models to the appendix (Sec.\ref{sec:model_variants}) for completion.

\subsection{Conditional complex-valued polynomial}
\label{sssec:complex_poly_method_two_variable} 
The aforementioned polynomial expansions rely on a single input variable, however often there are additional variables available, e.g., class-label information. In this case, we can design polynomial expansions from multiple input variables. Motivated by the real-valued multivariate analysis~\citep{chrysos2021conditional}, we focus on the case of two complex-valued inputs $\widetilde{\boldsymbol{x}}\in \mathbb{C}^{d}$ and $\widetilde{\boldsymbol{\psi}}\in \mathbb{C}^{d}$, but the expansions can be trivially extended to arbitrary number of inputs and also support different dimensions for the inputs.
Unlike \eqref{equ:3_3}, our goal turns to learn an $N^{\text{th}}$ degree polynomial expansion with an $o-$dimensional output $\boldsymbol{\widetilde{y}}$ with two inputs:
\begin{equation}
    \boldsymbol{\widetilde{y}} 
    = \sum_{n=1}^N \sum_{\gamma =1}^{n+1} \bigg(\boldsymbol{\mathcal{\widetilde{W}}}^{[n, \gamma ]} \prod_{j=2}^{\gamma } \times_{j} \widetilde{\boldsymbol{x}}  \prod_{\tau=\gamma +1}^{n+1} \times_{\tau} \widetilde{\boldsymbol{\psi}}\bigg) +\boldsymbol{\widetilde{\prodbias}}
    \label{eq:complex_poly_general_eq_2var_suppl}
\end{equation}  
Similar to the single-variable polynomial, \eqref{eq:complex_poly_general_eq_2var_suppl} could be
expressed to the following recursive relationship by applying coupled CP decomposition:
\begin{equation}
    \boldsymbol{\widetilde{y}}_{n} =  \Big(\boldsymbol{U}\matnot{n, I}^T \widetilde{\boldsymbol{x}} +  \boldsymbol{U}\matnot{n, II}^T \widetilde{\boldsymbol{\psi}}\Big) * \boldsymbol{\widetilde{y}}_{n-1}  + \boldsymbol{\widetilde{y}}_{n-1} 
    \label{eq:complex_poly_model1_rec}
\end{equation}
for $n=2,\ldots,N$ with $\boldsymbol{\widetilde{y}}_{1} = \boldsymbol{U}\matnot{1, I}^T \widetilde{\boldsymbol{x}} +  \boldsymbol{U}\matnot{1, II}^T \widetilde{\boldsymbol{\psi}}$ and
$\boldsymbol{\widetilde{y}}=\boldsymbol{\widetilde{H}} \boldsymbol{\widetilde{y}}_{N}+\boldsymbol{\widetilde{\prodbias}}$, where 
$ \boldsymbol{U}\matnot{n, I} 
\in 
\realnum^{d\times k},
\boldsymbol{U}\matnot{n, II} 
\in 
\realnum^{d\times k}.
$
Note that other decompositions discussed in Sec.\ref{sec:model_variants} can also be used.

\subsection{Adversarial audio generation}
\label{sec:methodaudiogeneration}
\begin{figure}[!ht]
\begin{minipage}[tb]{0.45\linewidth}
\centering
\includegraphics[width=\textwidth]{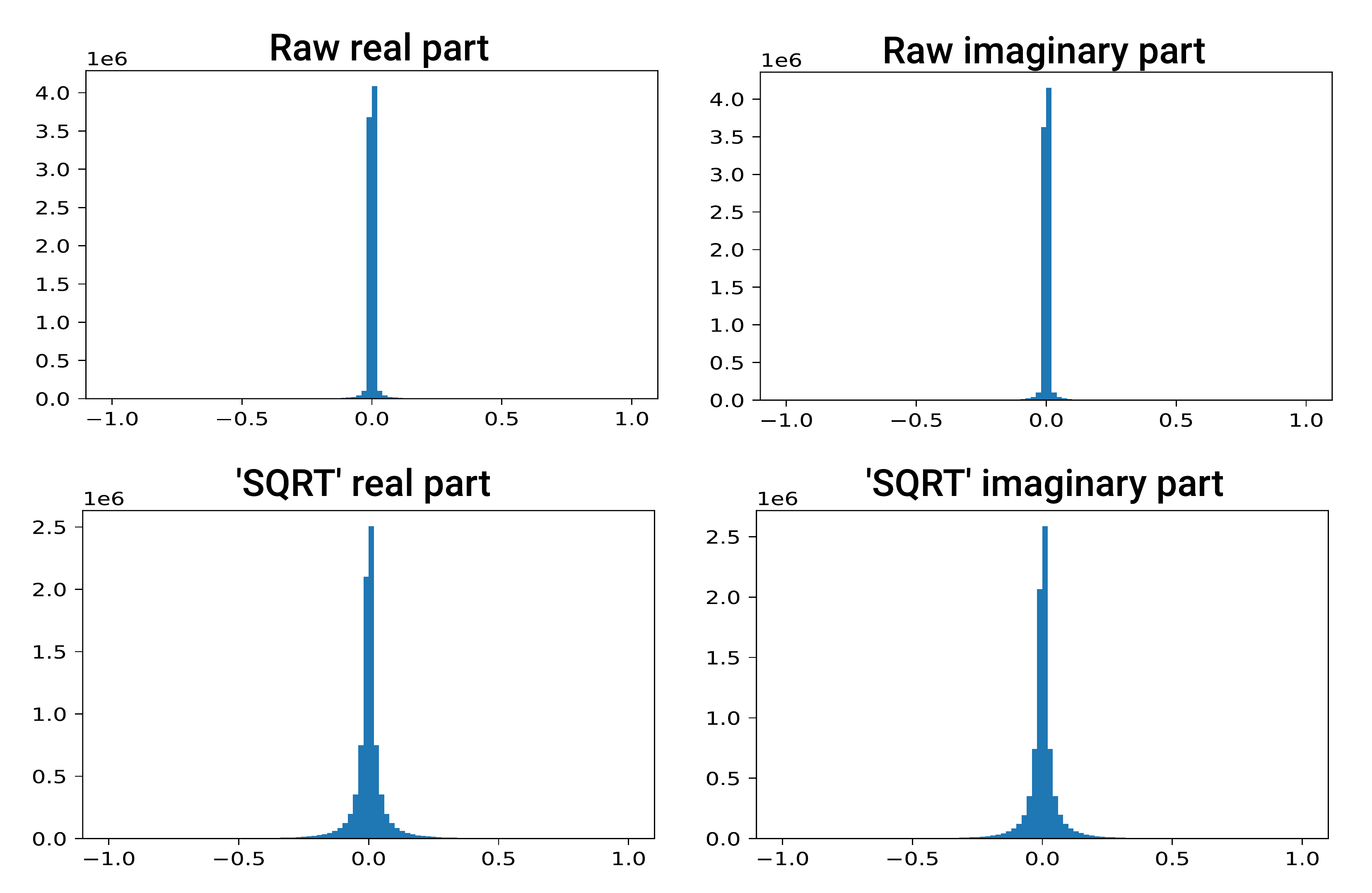}
\vspace{-5mm}
\caption{\textbf{Top}: The raw distribution of the real part and the imaginary part of STFT. \textbf{Bottom}: The distributions of the real part and the imaginary part of STFT with pre-processing techniques 'SQRT' become more even.}
\label{fig:distribution}
\end{minipage}
\hspace{0.5cm}
\begin{minipage}[tb]{0.45\linewidth}
\centering
\includegraphics[width=\linewidth]{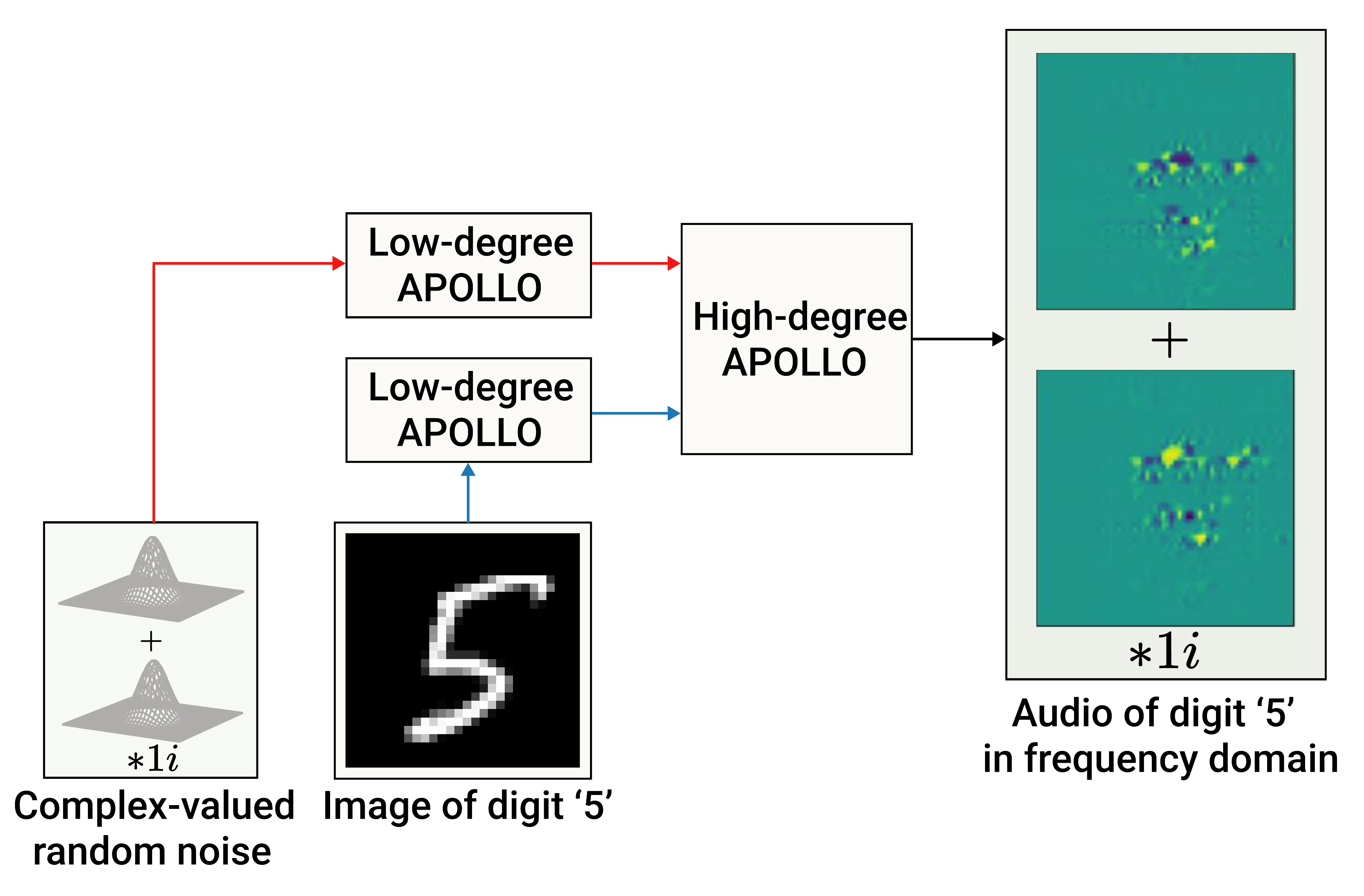} 
\caption{Generator used in image-to-speech experiments. 
A two-variable, high-degree \compoly{} is utilized to capture the correlations of the two input variables. 
}
\label{fig:image2audio}
\end{minipage}
\end{figure}

In the majority of our experimental validation we use GANs, where \compoly{} is chosen as the generator while the discriminator is a standard \resnet{}.
Wasserstein loss with gradient penalty is used as the criterion of GAN due to its stability and robustness~\cite{gulrajani2017improved}.
On unconditional audio generation, we implement the generator using single-variable models, e.g., the models in Sec.\ref{sec:model_comlex_valued_coefficients}, ~\ref{sec:model_real_valued_coefficients}. The generator receives 
a complex-valued noise and outputs the representation of audio in the frequency domain, as illustrated in Fig.\ref{fig:prodpoly_model_intro_schematic}.
Given an audio clip, we apply STFT and truncate the Nyquist bin to obtain the complex-valued representation. As visually depicted in Fig.\ref{fig:distribution}, the imbalanced distribution of the raw real part and the imaginary part of STFT is not well-suited for the Tanh activation of the generator. We introduce a series of pre-processing steps for training. Specifically, we first divide the real part and the imaginary part by its maximum value to limit the range between [-1,1] in order to match the output distribution of tanh activation function in the generator. 
In view of previous work that model the spectrum in log-scale, we take the square root of the absolute value of the real part (the imaginary part) and keep its sign.
This technique is denoted as 'SQRT' in Fig.\ref{fig:distribution}. Lastly, unless mentioned otherwise, we assume that ReLU activation are inserted after each degree, which result in a piece-wise polynomial expansion and tanh activation after the final degree.

When class label is available, we can also use a class-conditional generator, in which case we choose the two-input polynomial expansion.
Given a noise vector $\widetilde{\boldsymbol{x}} \in  \complexnum^{d_\text{1}}$ and one-hot label vector
$\widetilde{\boldsymbol{\psi}} \in  \complexnum^{d_\text{2}}$ with zero imaginary part,
the generator can be implemented based on \eqref{eq:complex_poly_model1_rec}, as shown in Fig.\ref{fig:prodpoly_model_conditional}.

Previous methods focus exclusively on a single modality, i.e., audio. However, as humans we perceive information using varying sources from the real-world. To
this end, we extend our APOLLO to multimodal generation (from image to audio). A schematic of the generator is visually depicted in Fig.\ref{fig:image2audio}.
Specifically,  we first use two low-degree APOLLOs for the random noise and the image, respectively. A high-degree conditional \compoly{}, proposed in Sec.\ref{sssec:complex_poly_method_two_variable}, is utilized to capture the correlations between the outputs of these two low-degree \compoly{}s and generate the complex-valued representation, e.g., STFT and CQT of the audio.

\textbf{Difference from \pinet{}}:
APOLLO differs substantially from $\Pi$-Nets in: a) The new decompositions that yield \eqref{eq:biasmodelcccp} and \eqref{eq:polygan_recursive_gen_third_order_lccp_final} are designed for reducing the number of parameters when extending to complex field and increasing the expressivity with the new bias term. b) We design architectures and technique for audio generation in Sec.\ref{sec:methodaudiogeneration} while \pinet{} is mostly focused on image-related tasks. c) \pinet{} have been used for a single variable input, while we also demonstrate experiments with two variables. e.g., conditional generation and multi-modal generation. Further discussion can be found in Sec.\ref{sec:differencefrompinet}. \section{Experiments}
We first conduct a series of experiments on audio generation to evaluate our framework. Then we further analyze the trained models. 
Lastly, we extend our model to additional audio-related tasks.
\label{sec:complex_poly_experiments}
\subsection{Comparison against adversarial-based models in audio generation}
\subsubsection{Unconditional audio generation}
\label{ssec:Unsupervised_audio_generation}
Our first experiment evaluates \compoly{} on unconditional audio generation, where the generator receives complex-valued noise and outputs the TF representation of audio, as illustrated in Fig.\ref{fig:prodpoly_model_intro_schematic} (appendix).
We conduct experiments on three datasets used in \citet{donahue2018adversarial}: (a) Speech Commands Zero Through Nine (SC09). (b) Piano dataset. (c) Drum dataset. Further details on the dataset, evaluation metrics, and experimental setup are offered in Sec.\ref{sec:dataset},~\ref{sec:metrics},~\ref{sec:detail_unsupervised_generation} respectively.
\begin{table*}[htb]
    \centering
    \caption{Comparison with adversarial methods on unconditional generation. Higher IS (lower FID, NDB, JSD) indicates better performance. The symbol '\# par’ abbreviates the number of parameters (in millions).~\compoly{} improves upon all the baselines in all metrics. Moreover, 
    \compoly-Small achieves similar performance with the baselines while reducing parameters by more than $87\%$.
     }
     \begin{tabular}{|c | c|c|c|c|c|} 
     \hline
     \multicolumn{6}{|c|}{Unconditional audio generation on SC09 dataset}\\ 
     \hline
     Model & IS ($\uparrow$) & FID ($\downarrow$)  & NDB ($\downarrow$)  & JSD ($\downarrow$) & \# par (M)\\
    \hline
     Real data& $8.01 
     $  &$0.50$ &$0.00
     $ & $0.011$&$\_$ \\
          \hline
     WaveGAN~\cite{donahue2018adversarial} 
     & $4.67
     $ & $41.60$&$ 16.00
     $ &$0.094$  & $36.5$\\
     SpecGAN~\cite{donahue2018adversarial}  
     & $6.03
     $ &$\_$ &$\_$ &$\_$ & $36.5$\\
    TiFGAN~\cite{marafioti2019adversarial}
    & $5.97$ &$26.70$ & $6.00
    $&$0.051$ &$42.4$\\
Mel-Spec GAN~\cite{Gunasekaran2020ImprovedSS} 
    & $5.76$ &$\_$ & $\_$&$\_$ &$\_$\\
    BigGAN~\cite{haque2020guided}
       & $6.17
       $ &$24.72$ & $\_$&$\_$ &$\_$ \\
          \pinet{}~\cite{chrysos2020p} 
          & ${6.59
          }$ & $13.01$&$4.40 
          $&0.048 &45.9 \\
\hline
\compoly, Small& $6.48
$ & $18.90$&$4.20
$ &$0.038$ & $4.6$\\
\compoly{}
&$\boldsymbol{
7.25
}$ &$\boldsymbol{8.15}$ &$\boldsymbol{3.20
}$ &$\boldsymbol{0.029}$& 64.1\\
\hline
\end{tabular}
\label{tab:poly_exper_audio_IS}
\end{table*}

\begin{figure}[!htb]
    \centering 
     \begin{subfigure}[tb]{0.49\textwidth}
         \centering \includegraphics[width=0.9\textwidth]{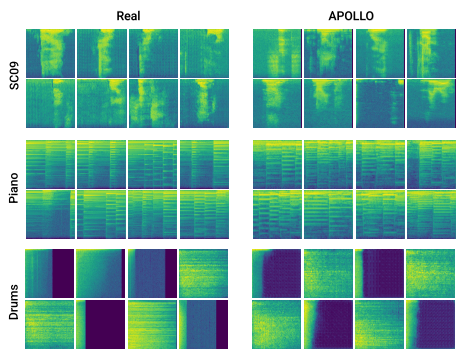}
         \caption{The log spectrums of the real samples and generated samples trained by our models on SC09, Piano, and Drum datasets. For each image, the horizontal (vertical) axis is along the time (frequency). The frequency increases with interval scale from 0 HZ (top) to 8000 HZ (bottom).}
     \label{fig:visualizespec}
     \end{subfigure}
     \hfill
     \begin{subfigure}[tb]{0.49\textwidth}
         \centering
     \includegraphics[width=0.9\textwidth]{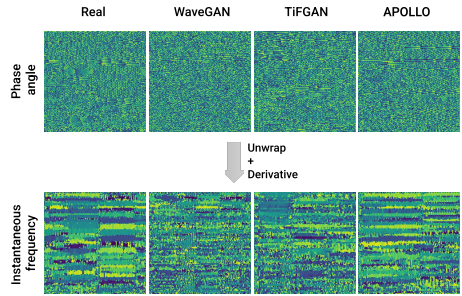}
         \caption{The phase information of STFT of the generated samples from different models trained on Piano dataset. Since it is hard to distinguish and compare with the phase angle, we unwrap the angle over
the $2\pi$ and take its derivative to obtain the instantaneous frequency. Notice that WaveGAN fails to generate realistic instantaneous frequency in most of frequency bins. The instantaneous frequency generated by our model is most consistent with the real one in terms of sharpness.}
    \label{fig:visualizeIF}
     \end{subfigure}
  \caption{This figure shows the the representation of the generated audios in the frequency domain.}
\end{figure}
\textbf{Results:} 
The log spectrums of audios synthesized by our model are presented in Fig.\ref{fig:visualizespec}.
Quantitative evaluations with adversarial-based models on SC09 dataset are reported in Table.\ref{tab:poly_exper_audio_IS}.
 \compoly{} obtains a large improvement upon all the baselines
 in Inception Score (IS)~\cite{NIPS2016_8a3363ab}, Frechet Inception Distance (FID)~\cite{NIPS2017_8a1d6947}, Number of Statistically-Different Bins (NDB), and Jensen-Shannon Divergence (JSD)~\cite{10.5555/3327345.3327486}. 
 The corresponding `Small' model performs similarly to the compared baselines while reducing parameters by more than $87\%$. 
Quantitative evaluations on Piano dataset are summarized in Table.\ref{tab:poly_exper_audio_Piano}, where we observe that \compoly{} improves upon the baselines in NDB and JSD while reducing parameters by more than $69\%$. 
Furthermore, we are interested in investigating whether \compoly{} could capture the phase information without explicitly modeling the phase in GANs or utilizing additional phase reconstruction technique. To this end, the phase angle and instantaneous frequency produced by different models trained on Piano dataset are visually depicted in Fig.\ref{fig:visualizeIF}.  We observe that WaveGAN fails to generate realistic instantaneous frequency in most of frequency bins. The instantaneous frequency generated by our model is the most consistent with the real one in terms of sharpness, which demonstrates that \compoly{} could better capture the phase information.

\begin{table}
\centering
\makebox[0pt][c]{\parbox{1\textwidth}{
    \begin{minipage}[!t]{0.52\linewidth}
    \setlength{\tabcolsep}{2pt}
    \caption{Quantitative evaluation on Piano dataset.
 \compoly{} outperforms the compared baselines by a large margin 
 while the small \compoly{} has $69\%$ fewer parameters than WaveGAN.}
     \begin{tabular}{|c | c|c|c|}
     \hline
     \multicolumn{4}{|c|}{Unconditional audio generation on Piano dataset}\\
     \hline
     Model & NDB ($\downarrow$) & JSD  ($\downarrow$)& \# par (M)\\
    \hline
     Real data& $0.00
     $ & $ 0.008$& $\_$\\
     \hline
     WaveGAN
     & $24.00
     $ & $ 0.547$&36.5\\
     TiFGAN
     & $17.60
     $ & $0.332$&$42.4$
     \\ \hline
          \compoly{}, Small& $13.20
          $ &  $0.270$&$\boldsymbol{11.3}$ \\
     \compoly{}& $\boldsymbol{8.80
     }$ &  $\boldsymbol{0.157}$&$42.7$ \\     \hline
     \end{tabular}
     \label{tab:poly_exper_audio_Piano} 
    \end{minipage}
    \hfill
    \begin{minipage}[!t]{0.46\linewidth}
    \setlength{\tabcolsep}{2pt}
    \caption{Quantitative Evaluation on conditional audio generation on SC09 dataset.
    \compoly{} improves upon BigGAN and Mel-Spec GAN by a considerable margin.
    }
     \begin{tabular}{|c | c|c|c|} 
     \hline
     \multicolumn{4}{|c|}{Conditional audio generation on SC09 dataset}\\ 
     \hline
     Model & IS ($\uparrow$)  & FID ($\downarrow$)& \# par (M) \\
    \hline
     Real data&  $8.01 
     $ & $0.50$ &$\_$\\
     \hline
     BigGAN      
     & $7.33
     $ & $ 24.40$ &$\_$\\ \hline
      Mel-Spec GAN
    & $7.64$ &$\_$ &\_\\
         \hline 
        \compoly{}& $\boldsymbol{7.73
         } $& $ \boldsymbol{6.31}$  &$61.7$\\  \hline
     \end{tabular}
     \label{tab:poly_exper_audio_sc09_con}
    \end{minipage}
}}
\end{table}
\subsubsection{Conditional audio generation}
\label{ssec:complex_conditional_audio_generation}
 In this section, we examine the proposed \compoly{} in conditional audio generation.
 Further details on the experimental setup are deferred to Sec.\ref{sec:detail_condition_generation}.
Regarding the dataset: (a) We first study the SC09 dataset, which has been used in the task of unconditional generation. The label is the class of digits.
(b) We also investigate a large music dataset, called NSynth, that consists 300,000 musical notes labelled with pitch, instrument, acoustic qualities, and velocity~\cite{nsynth2017}. Each sample lasts for 4 seconds. We use the same subset and the same test/train split as in \citet{engel2019gansynth}. The labels are the pitches ranging from MIDI 24 ($\sim$32Hz) to MIDI 84 ($\sim$1000Hz). 

\textbf{Result:} 
(a) The results in Table.\ref{tab:poly_exper_audio_sc09_con} demonstrate the improvement over conditional BigGAN~\cite{haque2020guided} and conditional Mel-Spec GAN~\cite{Gunasekaran2020ImprovedSS} on SC09 dataset. 
The visualization of the generated samples and the interpolation experiment are presented in  Sec.\ref{sec:detail_condition_generation} and Sec.\ref{ssec:complex_poly_interpolation} respectively.
(b) The result on Nsynth dataset are presented in Table.\ref{tab:poly_exper_audio_nsynth}. \compoly{}  improves upon GANSynth~\cite{engel2019gansynth} in terms of FID, NDB, JSD, the required samples for training, while having significantly less parameters.
\begin{table}[htb]
\setlength{\tabcolsep}{2pt}
    \centering
    \caption{Quantitative evaluation on conditional audio generation.     '\#sam’ abbreviates the total number of samples used during training (in millions).
    \compoly{} improves upon GANSynth in terms of FID, NDB, and JSD.
    It is rather remarkable that GANSynth is trained 
    with a batch size of 8 with 11 millions samples as reported in \citet{engel2019gansynth} while~\compoly{} is trained with the same batch size with only 0.48 millions samples. Furthermore,~\compoly{} has $24\%$ fewer parameters than GANSynth.
    }
     \begin{tabular}{|c |c|c|c|c|c|} 
     \hline
     \multicolumn{6}{|c|}{Conditional audio generation on Nsynth dataset}\\ 
     \hline
     Model &FID($\downarrow$) &NDB ($\downarrow$)  & JSD ($\downarrow$) &\#sam (M)&\# par (M)\\
    \hline
     Real data&   1.44&0.00&0.002&$\_$&$\_$\\
     \hline  
    GANSynth
     &3.91& 30.20&$ 0.362$ &11.00&14.1\\ \hline
     \compoly{}&
     $ \boldsymbol{1.98}$& $\boldsymbol{27.40} $& $\boldsymbol{0.298}$&$\boldsymbol{0.48}$&$\boldsymbol{10.6}$\\
      \hline
     \end{tabular}
     \label{tab:poly_exper_audio_nsynth}
\end{table}
\subsection{Comparison against non-adversarial models}
\label{ssec:Unsupervised_audio_generation_nonadversarial}
Previously, we have shown the comparison between our \compoly{} and other adversarial-based models with TF representation and have exhibited the excellent performance of \compoly{}.
As a compensation, in this section, we compare \compoly{} with non-adversarial models using waveform representation, e.g., SampleRNN~\citep{mehri2016samplernn}, WaveNet~\citep{oord2016wavenet}, DiffWave~\citep{kong2021diffwave}, SASHIMI~\citep{goel2022s}.
This experiment of unconditional generation is conducted on SC09 dataset. The dataset splitting and the details on evaluation metrics e.g., Modified Inception Score~\citep{gurumurthy2017deligan}, AM Score~\citep{zhou2018activation}, Inception Score (IS)~\citep{NIPS2016_8a3363ab}, Fréchet Inception Distance (FID)~\citep{NIPS2017_8a1d6947} are the same as in \citet{goel2022s}.
We provide the results in Table.\ref{tab:poly_exper_audio_IS_diffusion}, which shows that \compoly{} outperforms the baselines by a large margin.
\begin{table*}[!h]
    \centering
    \caption{Comparison with non-adversarial methods. Higher IS, MIS (lower FID, AM) indicate better performance.~\compoly{} largely improves upon the baselines.
    As a remark, previous experiment on SC09 use the same protocol as in 
    \citet{donahue2018adversarial} while in this experiment, the dataset splitting and the details on evaluation metrics are the same as in \citet{goel2022s}.}
     \begin{tabular}{|c | c|c|c|c|c|} 
     \hline
     \multicolumn{6}{|c|}{Unconditional audio generation on SC09 dataset}\\ 
     \hline
     Model & IS ($\uparrow$) & FID ($\downarrow$)  & MIS ($\uparrow$)  & AM ($\downarrow$) & \# par (M)\\
    \hline
     Real data& $8.33$  &$0.02$ &$257.6$ & $0.19$&$\_$ \\ \hline
    SampleRNN~\citep{mehri2016samplernn}& $1.71 $  &$8.96 $ &$3.02  $& $1.76 $&$35.0$ \\
        WaveNet~\citep{oord2016wavenet}& $2.27 $  &$5.08 $ &$5.80  $& $ 1.47$&$4.2$ \\
            DiffWave~\citep{kong2021diffwave}& $5.26 $  &$ 1.92$ &$51.21 $& $ 0.68 $&$24.1$ \\
                SASHIMI~\citep{goel2022s}& $5.94 $  &$ 1.42$ &$ 69.17$& $0.59 $&$23.0$ 
                \\
\hline
\compoly{}, Small& 
$6.03$ &$0.69$ &$77.05$ &$0.53$& $4.6$\\
\compoly{}&
$\boldsymbol{6.43}$ & $ \boldsymbol{0.45}$&$\boldsymbol{105.82}$ &$\boldsymbol{0.45}$ & $64.1 $\\
    \hline
\end{tabular}
\label{tab:poly_exper_audio_IS_diffusion}
\end{table*}
\subsection{Multimodal generation: Image-to-speech}
In this section, we access \compoly{} in multimodal generation in an image to audio experiment. 
In the experiment, a low-degree \compoly{}
is used for the complex-valued random noise. Since the pixels of image are real-valued, the corresponding low-degree polynomial is trivially chosen as models with real-valued coefficients.
Finally, a high-degree conditional \compoly{} with two input variables is chosen to capture the correlations and generate the STFT of the audio. 
We select SC09 as a source of digit audios and MNIST dataset \cite{lecun2010mnist} as a source of digit images. For each audio clip in the training set ($18620$ audio clips) of SC09, we align $2$ images with the same digit from the training set ($55000$ images) of MNIST without replacement. In total, we create $18620 \times 2$ audio-image pairs as the training set. We sample images from the testing set of MNIST to evaluate the model. The metrics reported in Table.\ref{tab:poly_exper_audio_multimodal}  is calculated by the same pretrained classifier used in Table.\ref{tab:poly_exper_audio_IS}. Further details on the experimental setup are deferred to
Sec.\ref{sec:detail_multimodal}.

\textbf{Results:} The results in Table.\ref{tab:poly_exper_audio_multimodal} indicate that the best model achieves $72\%$ accuracy. 
This experiment is more challenging than the corresponding class-conditional generation, owing to the different modality of the input-output pair, instead of the clean one-hot labels provided in the class-conditional generation. That explains the decrease in the score. 
\begin{table}[htb]
    \centering
    \setlength{\tabcolsep}{2pt}
    \caption{Quantitative evaluation on Image-to-speech generation on MNIST-SC09 dataset. '\#acc' abbreviates the categorial accuracy of the digit of the generated audio.
     }
     \begin{tabular}{|c | c|c|c|c|} 
     \hline
     \multicolumn{5}{|c|}{Image-to-speech generation on MNIST-SC09 dataset}\\ 
     \hline
     Model & IS ($\uparrow$)  & FID ($\downarrow$)& \#acc ($\uparrow$) & \# par (M)\\
    \hline
     Real data&  $8.01 
     $ & $0.50$ &$0.93$&\\
          \hline
 \compoly{}, Small& $5.75
 $& $26.5$&$0.68$ &$\boldsymbol{3.5}$\\ \hline
 \compoly{}& $\boldsymbol{6.90
 }$& $\boldsymbol{9.58}$&$\boldsymbol{0.72}$ &$46.1$\\ \hline
     \end{tabular}
     \label{tab:poly_exper_audio_multimodal}
\end{table}

\subsection{Further analysis}
\label{Sec:furtheranalysis}
We conduct further studies and comparisons on models trained on SC09 for unconditional generation.

\textbf{Inference speed.} From Table\ref{tab:inferencespeed}, we observe that WaveGAN has the highest inference speed since it synthesizes the audio directly in the time domain.
Even though \compoly{} can directly output the STFT without phase reconstruction, \compoly{} has an augmented inference time due to the complex operations, e.g., complex multiplication. The comparison with \pinet{} confirms our hypothesis, since they have a similar number of parameters but operate on the real field and they are faster than the corresponding \compoly{}. Nevertheless, when compared with models proposed specifically for audio generation, \compoly{} is still faster than TiFGAN, which requires phase reconstruction.
A future step for our model would be to further accelerate the complex multiplications, e.g., by implementing them directly in BLAS, instead of the high-level python operations.
On the other hand, non-adversarial models, e.g., DiffWave, result in the slowest inference speed due to the reverse process.

\textbf{Human study.}
We invite 25 volunteers
to assign an ordinal-scale score ($1$ to $5$) to each audio clip
based on the sound quality and perception. The qualitative results are summarized in Sec.\ref{sec:humanevaluation}. Our model
obtains the highest Mean Opinion Score (MOS)~\cite{Ribeiro2011CROWDMOSAA} with respect to the prior art.

\textbf{Ablation study.}
A thorough self-evaluation, e.g. interpolation of the inputs or empirical comparison of between different derivations, is deferred to Sec.
\ref{sec:ablation}
(Appendix) due to the constrained space.

\subsection{Beyond audio generation}
\label{sec:beyond_audio_generation}
Apart from audio generation, our model can also be adapted in other audio-related tasks. 
In the experiment of speech enhancement on VoiceBank-DEMAND dataset~\citep{thiemann2013diverse,veaux2013voice}, we demonstrate the advantage of \compoly{} over complex-valued neural networks and real-valued neural networks.
In additional, we conduct experiment on speech recognition with Speech Commands dataset~\citep{warden2018speech} and showcase the performance of our \compoly{}. Details can be found in Sec.\ref{sec:appendix_beyond_generation} of the appendix. \section{Conclusion}
\label{sec:complex_poly_conclusion}
In this work, we propose \compoly, a high degree complex-valued polynomial expansion for audio-related tasks. The parameters of the polynomial expansion are naturally expressed as high-order tensors, and we utilize standard tensor decompositions to reduce the parameters and implement the expansion with standard ML frameworks. Using different tensor decompositions results in diverse architectures with simple recursive formulations. The interested practitioner can easily derive custom architectures using complex-valued polynomial expansions by changing the factorizations behind the tensor decomposition as we illustrate. To validate the architectures, we conduct a thorough experimental validation in adversarial audio generation.
\compoly{} outperforms all the prior art by a large margin demonstrating the expressivity of the proposed complex-valued polynomial expansions. We believe this class of functions will be beneficial for synthesizing long audio tracks in the future. 
Furthermore, our experiments on conditional generation highlight the efficacy of \compoly{} on multimodal generation, where the extension to large-scale models, e.g., realistic text-to-speech translation, can be an interesting application for future work.

\section*{Acknowledgements}
\label{sec:acks}
This project has received funding from the European Research Council (ERC) under the European Union's Horizon 2020 research and innovation programme (grant agreement number 725594 - time-data). This project was partly supported by Zeiss. This work was supported by the Swiss National Science Foundation (SNSF) under  grant number 200021$\_$205011.
\bibliography{bibliography}
\bibliographystyle{plainnat}

\newpage
\appendix
\section*{Contents in the appendix}
\label{sec:complex_poly_content_appendix}
The appendix contains further details and derivations of the \compoly{}. Specifically, a detailed notation is developed in Sec~\ref{sec:complex_poly_detailed_notation}. The variants of \compoly{} are studied in Sec~\ref{sec:complex_valued_variant_appendix},~\ref{sec:real_valued_variant_appendix}.
The differences from \pinet{} are explored in Sec~\ref{sec:differencefrompinet}. In Sec~\ref{sec:model_derivation}, the derivations and the proofs in the main body are further elaborated. 
Regarding experiments, the evaluation metrics of audio generation are presented in Sec~\ref{sec:metrics}.
Further details of the experimental setup on unconditional audio generation, conditional audio generation, and multimodal generation are developed in Sec~\ref{sec:detail_unsupervised_generation},~\ref{sec:detail_condition_generation},~\ref{sec:detail_multimodal}, respectively. Ablation studies on our models are included in Sec~\ref{sec:ablation}.
Additional analyses for the inference speed and human study are discussed in Sec~\ref{sec:appendixinferencespeed},~\ref{sec:humanevaluation}.
Experiments on speech recognition and speech enhancement are included in Sec~\ref{sec:appendix_beyond_generation}.
Lastly, the societal impact is described in Sec~\ref{sec:complex_polynomial_societal_impact}. 

\section{Detailed notation}
\label{sec:complex_poly_detailed_notation}
\textbf{Symbols of variables}: As a reminder, real-valued matrices (vectors) are symbolized by uppercase (lowercase) boldface letters, e.g., $\boldsymbol{Y}$, $\boldsymbol{y}$.
Tensors are the multidimensional equivalent
of matrices.
Real-valued tensors are symbolized by calligraphic letters, e.g., $\boldsymbol{\mathcal{Y}}$.
All complex-valued variables are symbolized with wide tilde, e.g., $\boldsymbol{\widetilde{y}}$, $\boldsymbol{\widetilde{Y}}$, $\widetilde{\boldsymbol{\mathcal{Y}}}$. The real part (the imaginary part) of a complex-valued variable is accessed by $\mathfrak{Re}$ ($\mathfrak{Im}$).

\textbf{Complex-valued transpose}: 
In this paper, all transposes for complex-valued matrices are trivial transposes (the same as in real-valued field) without calculating complex conjugate.

\textbf{Matrix products}: The \textit{Khatri-Rao} product of two matrices $\boldsymbol{\widetilde{A}} \in \complexnum^{I \times N}$
and $\boldsymbol{\widetilde{C}} \in \complexnum^{J \times N}$ is
denoted by $\boldsymbol{\widetilde{A}} \odot \boldsymbol{\widetilde{C}}$.
The \textit{Khatri-Rao} product of multiple matrices $\{\boldsymbol{\widetilde{A}}\matnot{m} \in \complexnum^{I_m \times N} \}_{m=1}^M$ is denoted by $\boldsymbol{\widetilde{A}}\matnot{1} \odot \boldsymbol{\widetilde{A}}\matnot{2} \odot  \cdots \odot  \boldsymbol{\widetilde{A}}\matnot{M} \doteq  \bigodot_{m=1}^M \boldsymbol{\widetilde{A}}\matnot{m}$.
The \textit{Hadamard} product of  $\boldsymbol{\widetilde{A}}, \boldsymbol{\widetilde{C}} \in \complexnum^{I \times N}$ is denoted by $\boldsymbol{\widetilde{A}} * \boldsymbol{\widetilde{C}}$, where the $(i, j)$ element is equal to ${a}_{(i, j)} {b}_{(i, j)}$. 

\textbf{Tensors}: Consider an $M^{\text {th}}$ order tensor: $\boldsymbol{\mathcal{\widetilde{Y}}} \in \mathbb{C}^{J_{1} \times J_{2} \times \cdots \times J_{m-1} \times J_{m} \times J_{m+1} \times \cdots \times J_{M}}$. We address each element by $(\boldsymbol{\mathcal{\widetilde{Y}}})_{j_{1}, j_{2}, \ldots, j_{M}} \doteq y_{j_{1}, j_{2}, \ldots, j_{M}}.$
The \textit{mode-$m$} vector product between the tensor $\boldsymbol{\mathcal{\widetilde{Y}}}$ and a vector $\widetilde{\boldsymbol{u}} \in \mathbb{C}^{J_{m}}$ is denoted by $\boldsymbol{\mathcal{\widetilde{Y}}} \times_{m} \widetilde{\boldsymbol{u}} \in \mathbb{C}^{J_{1} \times J_{2} \times \cdots \times J_{m-1} \times J_{m+1} \times \cdots \times J_{M}}$, yielding an $(M\minus1)^{\text {th}}$ order tensor:
$$
\left(\boldsymbol{\mathcal{\widetilde{Y}}} \times_{m} \widetilde{\boldsymbol{u}}\right)_{j_{1}, \ldots, j_{m-1}, j_{m+1}, \ldots, j_{M}}=\sum_{j_{m}=1}^{J_{m}} \widetilde{y}_{j_{1}, j_{2}, \ldots, j_{M}} \widetilde{u}_{j_{m}}.
$$
The \textit{mode-$m$} vector product of a complex-valued tensor and multiple complex-valued vectors is denoted as:
$$\widetilde{\boldsymbol{\mathcal{Y}}} \times_{1} \widetilde{\boldsymbol{u}}^{(1)} \times_{2} \widetilde{\boldsymbol{u}}^{(2)} \times_{3} \cdots \times_{M} \widetilde{\boldsymbol{u}}^{(M)} =\widetilde{\boldsymbol{\mathcal{Y}}} \prod_{m=1}^{M} \times_{m} \widetilde{\boldsymbol{u}}^{(m)}.$$
As a generalization of \textit{CANDECOMP/PARAFAC (CP) decomposition}~\citep{kolda2009tensor} in complex field,  complex-valued tensor could be decomposed into a sum of component rank-one tensors. The rank-$R$ CP decomposition of an $M^{\text{th}}$-order tensor $\widetilde{\boldsymbol{\mathcal{Y}}}$ is denoted by:
 \begin{equation}\label{E:CP}
\widetilde{\boldsymbol{\mathcal{Y}}} \doteq [\![ \boldsymbol{\widetilde{U}}\matnot{1}, \boldsymbol{\widetilde{U}}\matnot{2}, \ldots, \boldsymbol{\widetilde{U}}\matnot{M}  ]\!] =  \sum_{r=1}^R \boldsymbol{\widetilde{U}}_r^{(1)}  \circ \boldsymbol{\widetilde{U}}_r^{(2)}  \circ \cdots \circ \boldsymbol{\widetilde{U}}_r^{(M)},
\end{equation}
where $\circ$ is the vector outer product. The factor matrices $\big\{ \boldsymbol{\widetilde{U}}\matnot{m} = [\boldsymbol{\widetilde{u}}_1^{(m)},\boldsymbol{\widetilde{u}}_2^{(m)}, \cdots, \boldsymbol{\widetilde{u}}_R^{(m)} ] \in \mathbb{C}^{I_m \times R} \big\}_{m=1}^{M}$ collect 
the vectors from the rank-one components. 
Particularly, we can express the CP decomposition in matrix form:
 \begin{equation}
 \label{eq:polygan_cp_unfolding}
\boldsymbol{\widetilde{W}}_{(1)}  
\doteq \boldsymbol{\widetilde{U}}\matnot{1} \bigg( \bigodot_{m = M}^{2} \boldsymbol{\widetilde{U}}\matnot{m}\bigg)^T.
\end{equation}
\begin{lemma}
Given the sets of real-valued matrices $\{\boldsymbol{A}_{\nu} \in \mathbb{R}^{I_{\nu} \times K} \}_{\nu=1}^N$  and $\{\boldsymbol{C}_{\nu} \in \mathbb{R}^{I_{\nu} \times L} \}_{\nu=1}^N$, the following
equality holds:
\begin{equation}
    (\bigodot_{\nu=1}^N \boldsymbol{A}_{\nu})^T \cdot (\bigodot_{\nu=1}^N \boldsymbol{C}_{\nu}) = (\boldsymbol{A}_1^T \cdot \boldsymbol{C}_1) * \ldots * (\boldsymbol{A}_N^T \cdot \boldsymbol{C}_N).
\end{equation}
\label{lemma:polygan_lemma_hadamard_kr1}
\end{lemma}
The proof can be found in the appendix of \citet{chrysos2019polygan}. 
\begin{lemma}
 Given the sets of complex-valued matrices $\{\boldsymbol{\widetilde{A}}_{\nu} \in \mathbb{C}^{I_{\nu} \times K} \}_{\nu=1}^N$  and $\{\boldsymbol{\widetilde{C}}_{\nu} \in \mathbb{C}^{I_{\nu} \times L} \}_{\nu=1}^N$, the following
equality holds:
\begin{equation}
    (\bigodot_{\nu=1}^N \boldsymbol{\widetilde{A}}_{\nu})^T \cdot (\bigodot_{\nu=1}^N \boldsymbol{\widetilde{C}}_{\nu}) = (\boldsymbol{\widetilde{A}}_1^T \cdot \boldsymbol{\widetilde{C}}_1) * \ldots * (\boldsymbol{\widetilde{A}}_N^T \cdot \boldsymbol{\widetilde{C}}_N).
\end{equation}
\label{lemma:polygan_lemma_hadamard_kr2}
\end{lemma}
The proof is the same as that of Lemma \ref{lemma:polygan_lemma_hadamard_kr1} except the field ($\realnum$ or $\complexnum$) of the matrices.

\section{Model variants}
\label{sec:model_variants}
\begin{figure*}[!h]
    \centering
    \includegraphics[width=1\linewidth]{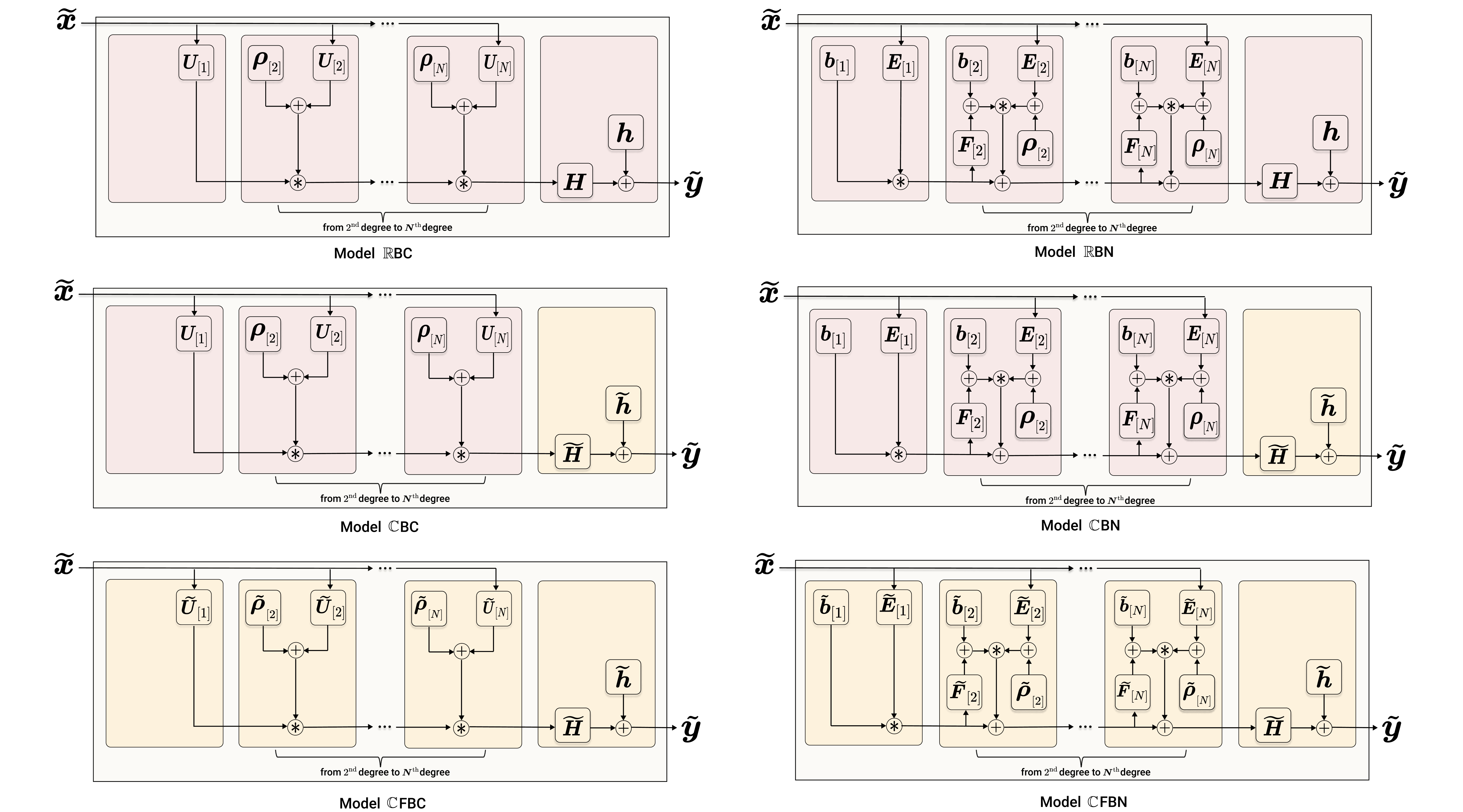}
\caption{Schematic summary of our proposed models. 
All learnable parameters inside pink blocks (yellow blocks) are real-valued (complex-valued).
We observe that \biasmodelmixccp{} (\biasmodelmixncp{}) reduces the number of parameters of \biasmodelcccp{} (\biasmodelcncp{})  by changing the field ($\complexnum{}\to\,\realnum{}$) of the parameters from 1$^{\text{st}}$ degree to $N^{\text{th}}$ degree.
} 
\label{fig:prodpoly_model_intro_schematic_all}
\end{figure*}

This section introduces different variants of our proposed framework.
The dimension and the field ($\realnum$ or $\complexnum$) of learnable parameters in all proposed models are summarized in Table \ref{tbl:prodpoly_primary_symbolsalll} and the corresponding schematics are depicted in Fig~\ref{fig:prodpoly_model_intro_schematic_all}.
In the main body, we have already provided three models, below we complement those with the remaining models.
 \begin{table*}[h]
\caption{Table with detailed symbol reference per model.}
\label{tbl:prodpoly_primary_symbolsalll}
\centering
\begin{tabular}{|c | c | c|}
\toprule
Symbol 	& Dimension(s) 		&	Definition \\
\midrule

$\boldsymbol{H}, 
\boldsymbol{\prodbias}, 
\boldsymbol{U}\matnot{n},\bm{\rho}\matnot{n}$
& 
$\realnum^{o\times k}, 
\realnum^{o},
\realnum^{d\times k},
\realnum^{k}
$	
&	Parameters in \biasmodelrccp{} \\

$\boldsymbol{H}, 
\boldsymbol{\prodbias},
\boldsymbol{E}\matnot{n},
\boldsymbol{F}\matnot{n},
\boldsymbol{b}\matnot{n},
\bm{\rho}\matnot{n}
$
& $\realnum^{o\times k}, 
\realnum^{o},
\realnum^{d\times k}, 
\realnum^{k\times k}, 
\realnum^{k},
\realnum^{k}$
&	Parameters in \biasmodelrncp{} \\
\hline

$\boldsymbol{\widetilde{H}}, 
\boldsymbol{\widetilde{\prodbias}}, 
\boldsymbol{U}\matnot{n},
\bm{\rho}\matnot{n}
$
& 
$\complexnum^{o\times k}, 
\complexnum^{o},
\realnum^{d\times k},
\realnum^{k}$	
&	Parameters in \biasmodelmixccp{} \\

$\boldsymbol{\widetilde{H}}, 
\boldsymbol{\widetilde{\prodbias}},
\boldsymbol{E}\matnot{n}, 
\boldsymbol{F}\matnot{n},
\boldsymbol{b}\matnot{n},
\bm{\rho}\matnot{n}
$
& $\complexnum^{o\times k}, 
\complexnum^{o},
\realnum^{d\times k}, 
\realnum^{k\times k}, 
\realnum^{k},
\realnum^{k}$
&	Parameters in \biasmodelmixncp{} \\

 \hline

$\boldsymbol{\widetilde{H}}, 
\boldsymbol{\widetilde{\prodbias}},
\boldsymbol{\widetilde{U}}\matnot{n},\bm{\widetilde{\rho}}\matnot{n}$
& 
$\complexnum^{o\times k},
\complexnum^{o},
\complexnum^{d\times k},
\complexnum^{k}
$
&	Parameters in \biasmodelcccp{} \\

$\boldsymbol{\widetilde{H}}, 
\boldsymbol{\widetilde{\prodbias}},
\boldsymbol{\widetilde{E}}\matnot{n}, 
\boldsymbol{\widetilde{F}}\matnot{n},
\boldsymbol{\widetilde{b}}\matnot{n}
,\bm{\widetilde{\rho}}\matnot{n}$
& $\complexnum^{o\times k}, 
\complexnum^{o},
\complexnum^{d\times k}, 
\complexnum^{k\times k}, 
\complexnum^{k},
\complexnum^{k}$
&	Parameters in \biasmodelcncp{} \\
\hline
\end{tabular}
\end{table*}

 \subsection{\compoly{}  with complex-valued coefficients}
 \label{sec:complex_valued_variant_appendix}
 In the main body, we have introduced fully coupled decomposition with complex-valued coefficients. Below, we provide its corresponding nested coupled decomposition variants.
 
\textbf{\biasmodelmixncp{}~(Nested decomposition with bias)}
Unlike \biasmodelmixccp{}, we apply a joint hierarchical decomposition instead of separating the interactions between different degrees. The learnable hyper-parameters 
$\boldsymbol{\beta}_{[n]} \in \mathbb{R}^{\omega}$ for $n=1, \ldots, N$, are introduced as scaling factors and \eqref{equ:3_3} becomes: 
\begin{equation}
    \boldsymbol{\widetilde{y}} = \sum_{n=1}^N \bigg(\boldsymbol{\mathcal{\widetilde{W}}}^{[n]} \times_2 \boldsymbol{\beta}\matnot{N+1-n} \prod_{j=3}^{n+2} \times_{j} \boldsymbol{\widetilde{x}}\bigg) + \boldsymbol{\widetilde{\prodbias}}.
    \label{eq:prodpoly_general_polynomial_with_lightenmodel}
\end{equation}
The interactions between different degrees are learned through a joint hierarchical decomposition on the polynomial parameters.
\begin{itemize}
    \vspace{-1.1em}
    \setlength\itemsep{-0.3em}
    \item First-degree parameters : $\boldsymbol{\widetilde{W}}^{[1]}_{(1)} = \boldsymbol{\widetilde{H}} (\boldsymbol{E}\matnot{3} \odot \boldsymbol{B}\matnot{3})^T$.
    \item Second-degree parameters:
    $\boldsymbol{\widetilde{W}}^{[2]}_{(1)} = \boldsymbol{\widetilde{H}} \bigg\{\boldsymbol{E}\matnot{3} \odot \bigg[\Big(\boldsymbol{E}\matnot{2} \odot \boldsymbol{B}\matnot{2}\Big) \boldsymbol{F}\matnot{3}\bigg]\bigg\}^T$.
    \item Third-degree parameters:
    $\boldsymbol{\widetilde{W}}^{[3]}_{(1)} = \boldsymbol{\widetilde{H}} \bigg\{\boldsymbol{E}\matnot{3} \odot \bigg[\bigg(\boldsymbol{E}\matnot{2} \odot \Big\{\Big(\boldsymbol{E}\matnot{1} \odot \boldsymbol{B}\matnot{1}\Big) \boldsymbol{F}\matnot{2}\Big\} \bigg)\boldsymbol{F}\matnot{3} \bigg]\bigg\}^T$,
\end{itemize}
where $\bm{\widetilde{H}} \in  \complexnum^{o \times k}$, 
$\bm{{E}}\matnot{n} \in  \realnum^{d \times k}$
$\bm{{B}}\matnot{n} \in  \realnum^{w \times k}$
$\bm{{F}}\matnot{n} \in  \realnum^{k \times k}$ for $n=1, \ldots, N$. 
Similarly, 
we can obtain the recursive form for $N^{\text{th}}$ degree expansion: 
\begin{equation*}
\boldsymbol{\widetilde{y}}_{n} = \left( {\boldsymbol{E}\matnot{n}^T\widetilde{\boldsymbol{x}}} \right) * \left( {\boldsymbol{F}\matnot{n}^T\boldsymbol{\widetilde{y}}_{n-1} +
\boldsymbol{b}\matnot{n}}
\right),
\end{equation*}
for $n=2, \ldots, N$ with $\boldsymbol{\widetilde{y}}_{1}=(\boldsymbol{E}_{[1]}^{T} \widetilde{\boldsymbol{x}}) *\left(\boldsymbol{b}_{[1]}\right)$, $\boldsymbol{\widetilde{y}}=$
$\boldsymbol{\widetilde{H}} \boldsymbol{\widetilde{y}}_{N}+\boldsymbol{\widetilde{\prodbias}}$, where we denote by $\boldsymbol{b}\matnot{n}= \boldsymbol{B}\matnot{n}^T\boldsymbol{\beta}\matnot{n}$ for $n=1, \ldots, N$. 
 Motivated by the skip connections in ResNet \cite{7780459}, we embed a shortcut connection into the relationship. In addition, we add a bias term $\bm{\rho}\matnot{n} \in 
\realnum^{k}$, which is similar to \biasmodelmixccp{}.
 The final recursive relationship:
\begin{equation}
\boldsymbol{\widetilde{y}}_{n} = \left( {\boldsymbol{E}\matnot{n}^T\widetilde{\boldsymbol{x}}}
+\bm{\rho}\matnot{n}
\right) * \left( {\boldsymbol{F}\matnot{n}^T\boldsymbol{\widetilde{y}}_{n-1} +\boldsymbol{b}\matnot{n}} \right) +{\boldsymbol{\widetilde{y}}}_{n-1},
\end{equation}
for $n=2, \ldots, N$ with $\boldsymbol{\widetilde{y}}_{1}=(\boldsymbol{E}_{[1]}^{T} \widetilde{\boldsymbol{x}}) *\left(
\boldsymbol{b}_{[1]}
\right)$ and $\boldsymbol{\widetilde{y}}=$
$\boldsymbol{\widetilde{H}} \boldsymbol{\widetilde{y}}_{N}+\boldsymbol{\widetilde{\prodbias}}.$
Note that all learnable parameters in all degrees are real-valued except in the highest degree.

\textbf{\biasmodelcncp{}~(Fully nested decomposition with bias)
} When all the parameters in \biasmodelmixncp{} are complex-valued, we have the following recursive relationship:
\begin{equation}
\boldsymbol{\widetilde{y}}_{n} = \left( {\boldsymbol{\widetilde{E}}\matnot{n}^T\widetilde{\boldsymbol{x}}}
+\bm{\widetilde{\rho}}
\matnot{n}
\right) * \left( {\boldsymbol{\widetilde{F}}\matnot{n}^T\boldsymbol{\widetilde{y}}_{n-1} +\boldsymbol{\widetilde{b}}\matnot{n}} \right) +{\boldsymbol{\widetilde{y}}}_{n-1},
\end{equation}
for $n=2, \ldots, N$ with $\boldsymbol{\widetilde{y}}_{1}=(\boldsymbol{\widetilde{E}}_{[1]}^{T} \widetilde{\boldsymbol{x}}) *\left(\boldsymbol{\widetilde{b}}_{[1]}\right)$, $\boldsymbol{\widetilde{y}}=$
$\boldsymbol{\widetilde{H}} \boldsymbol{\widetilde{y}}_{N}+\boldsymbol{\widetilde{\prodbias}}$, where we denote by $\boldsymbol{\widetilde{b}}\matnot{n}= \boldsymbol{\widetilde{B}}\matnot{n}^T\boldsymbol{\widetilde{\beta}}\matnot{n}$ for $n=1, \ldots, N$.

\subsection{\compoly{}  with real-valued coefficients}
\label{sec:real_valued_variant_appendix}
The models derived in the main body assume complex-valued coefficients.
However we can replace the coefficients in \eqref{equ:3_3} with its real-valued counterpart and obtain the following polynomial:
  \begin{equation}
\boldsymbol{\widetilde{Y}}=\sum_{n=1}^{N}\left(\boldsymbol{\mathcal{W}}^{[n]} \prod_{j=2}^{n+1} \times{ }_{j} \widetilde{\boldsymbol{x}}\right)+\boldsymbol{\prodbias},
\label{equ:3_2}
\end{equation}
where  $\left\{\boldsymbol{\mathcal{W}}^{[n]} \in \mathbb{R}^{o \times\prod_{m=1}^{n}{\times}_{m} d} \right\}_{n=1}^{N}$ and $\boldsymbol{\prodbias} \in \mathbb{R}^{o}$ are learnable parameters.
Similarly, we can apply different decomposition techniques to reduce the parameters and obtain the recursive relationship between different degrees. 
Since the only difference from the models given in the main paper is the field ($\realnum{}$ or $\complexnum{}$) of the learnable parameters, we skip the derivation and the recursive relationship. If we convert all the parameters in \biasmodelmixccp{} from complex-valued to real-valued, we can obtain \biasmodelrccp{}.
If we convert all the parameters in \biasmodelmixncp{} from complex-valued to real-valued, we can obtain \biasmodelrncp{}.

\subsection{In-depth difference from real-valued polynomial networks}
\label{sec:differencefrompinet}
The family of \pinet{} bears resemblance with the proposed \compoly, however those two differ substantially in the following ways: 
\begin{itemize}
    \item The family of \pinet{} was designed for real-valued inputs and outputs with all the learnable parameters being real-valued as well. On the contrary, the proposed \compoly{} enables using complex-valued inputs or outputs. 
    \item The motivation for designing the two models differs: in \compoly{} the goal is to construct an efficient method for time-frequency representations frequently met in audio-related tasks. On the contrary, \pinet{} is (mostly) focused on image-related tasks or non-euclidean meshes. 
     In Sec~\ref{sec:methodaudiogeneration}, we design several architectures and technique to adapt our decomposition to audio generation. As shown in Table~\ref{tab:poly_exper_audio_IS}, our APOLLO greatly outperforms \pinet{}, which are trivially adapted to audio generation.

    \item Beyond the technical tensor decompositions, a significant contribution in our case is to highlight which decompositions work well in the complex field. Even though there are several combinations of recursive formulations that could be designed, we showcase how to construct efficient recursive formulations in a principled way. 
    \item \pinet{} express a polynomial expansion of one variable, which presents an issue in case of multiple input variables being available. Particularly, in the case of the variables representing different type of data, e.g., vectors and tensors, it might be more convenient to not concatenate them. This was similarly reported in \citet{chrysos2021conditional}, and we follow that model in conditional generation. Concretely, in the image to audio experiment, \pinet{} needs to vectorize the input image to concatenate it with the noise, which would destroy the spatial correlations while in APOLLO, we utilize low-degree polynomial to capture the representation and then use a high-degree polynomial to learn the correlation. 
\end{itemize}
\begin{figure*}[!]
    \centering
    \includegraphics[width=1\linewidth]{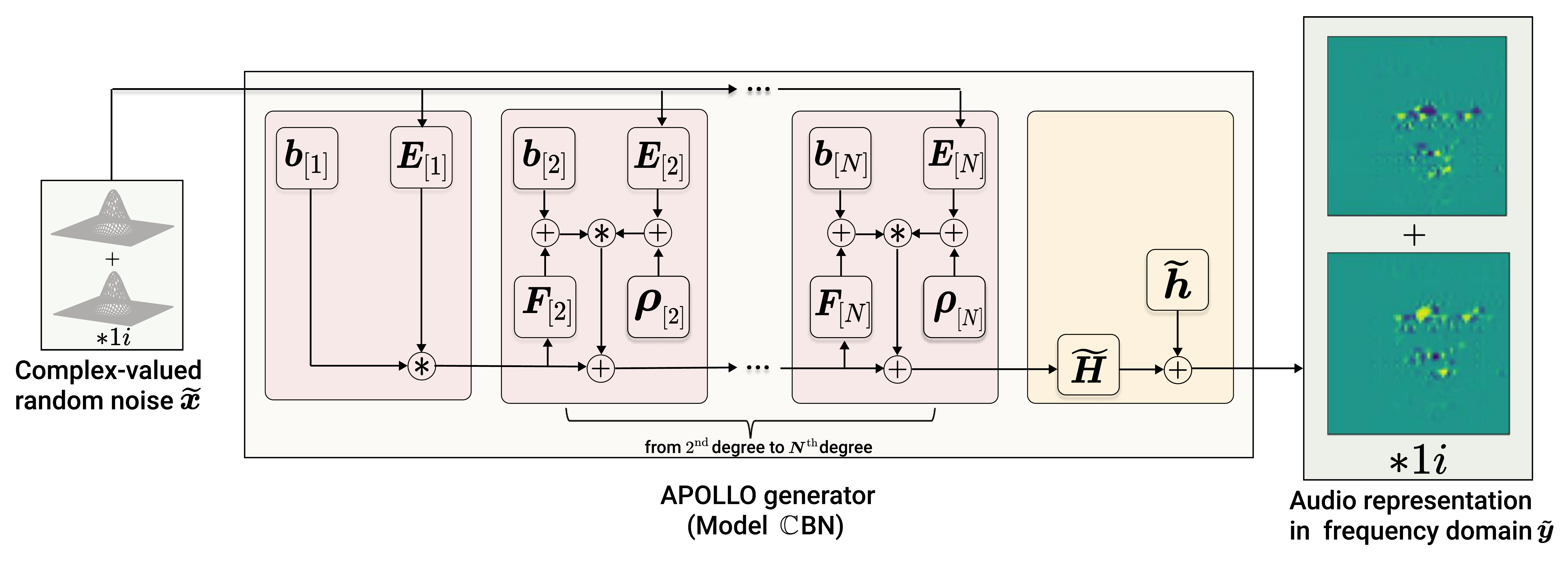}\vspace{-3mm}
\caption{Schematic of the unconditional \compoly{} generator, where the complex-valued output is a polynomial of the complex-valued input. The input of the generator is the complex-valued noise and the output is the representation of audio in the frequency domain (e.g., STFT, CQT). 
All learnable parameters inside the pink blocks (yellow blocks) are real-valued (complex-valued).}
\label{fig:prodpoly_model_intro_schematic}
\end{figure*}

\begin{figure*}[!h]
    \centering
    \includegraphics[width=0.9\linewidth]{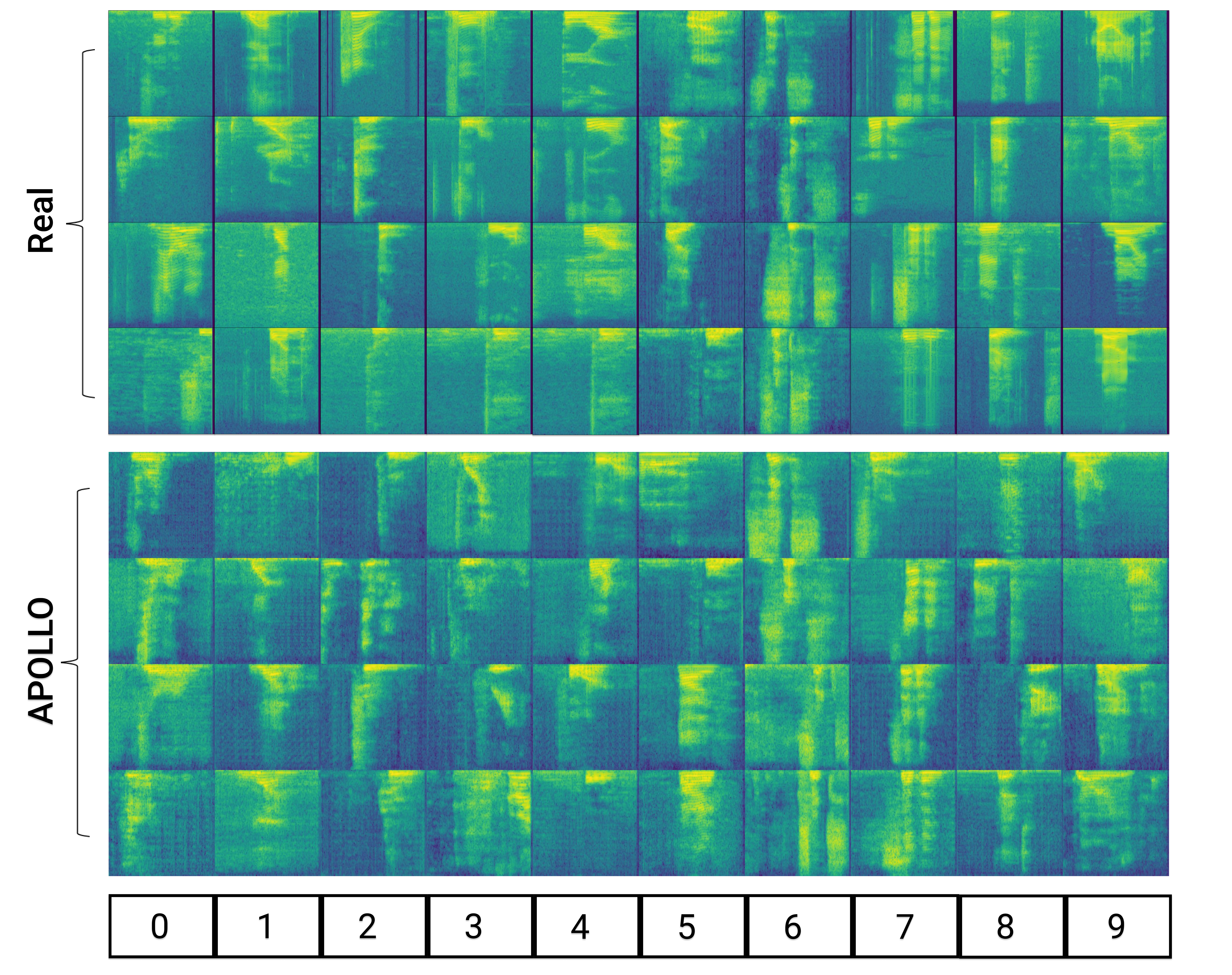}
\caption{The log spectrum of the real samples and the samples generated by \compoly{} trained on SC09 dataset for class-conditional generation. Each column indicates the category from 0 to 9. Our model is able to produce realistic log spectrum in the case of conditional generation.}
\label{fig:prodpoly_model_conditional_sampels}
\end{figure*}

\section{Model derivations}
\label{sec:model_derivation}
\subsection{Derivations for~\modelcccp}
\label{ssec:sup_for_cccp}
By leveraging the factorizations, i.e.,
\begin{itemize} 
    \vspace{-1.1em}
    \setlength\itemsep{-0.3em}
    \item First degree parameters: $\boldsymbol{\widetilde{W}}^{[1]} = \boldsymbol{\widetilde{H}}\boldsymbol{\widetilde{U}}\matnot{1}^T$.
    \item Second degree parameters: $\boldsymbol{\widetilde{W}}^{[2]}_{(1)} = \boldsymbol{\widetilde{H}}(\boldsymbol{\widetilde{U}}\matnot{3} \odot \boldsymbol{\widetilde{U}}\matnot{1})^T + \boldsymbol{\widetilde{H}}(\boldsymbol{\widetilde{U}}\matnot{2} \odot \boldsymbol{\widetilde{U}}\matnot{1})^T$.
    \item Third degree parameters: $\boldsymbol{\widetilde{W}}^{[3]}_{(1)} = \boldsymbol{\widetilde{H}}(\boldsymbol{\widetilde{U}}\matnot{3} \odot \boldsymbol{\widetilde{U}}\matnot{2} \odot \boldsymbol{\widetilde{U}}\matnot{1})^T $,
\end{itemize}
the third degree expansion of \eqref{equ:3_3} is expressed as:
\vspace{-2pt}
\begin{equation}
\begin{split}
     \boldsymbol{\widetilde{y}}&= \boldsymbol{\widetilde{\prodbias}} + \boldsymbol{\widetilde{H}}\boldsymbol{\widetilde{U}}\matnot{1}^T\widetilde{\boldsymbol{x}} + \boldsymbol{\widetilde{H}}\Big(\boldsymbol{\widetilde{U}}\matnot{3} \odot \boldsymbol{\widetilde{U}}\matnot{1}\Big)^T(\widetilde{\boldsymbol{x}}\odot \widetilde{\boldsymbol{x}}) + \boldsymbol{\widetilde{H}}\Big(\boldsymbol{\widetilde{U}}\matnot{2} \odot \boldsymbol{\widetilde{U}}\matnot{1}\Big)^T(\widetilde{\boldsymbol{x}}\odot\widetilde{\boldsymbol{x}})  \\
    & +\boldsymbol{\widetilde{H}}\Big(\boldsymbol{\widetilde{U}}\matnot{3} \odot \boldsymbol{\widetilde{U}}\matnot{2} \odot \boldsymbol{\widetilde{U}}\matnot{1}\Big)^T(\widetilde{\boldsymbol{x}} \odot \widetilde{\boldsymbol{x}} \odot \widetilde{\boldsymbol{x}}).
\label{eq:polygan_recursive_gen_third_order_cccp}
\end{split}
\end{equation}
    Applying Lemma \ref{lemma:polygan_lemma_hadamard_kr2} on (\ref{eq:polygan_recursive_gen_third_order_cccp}), we obtain:
    \begin{equation}
    \begin{split}
    &
    \boldsymbol{\widetilde{y}}=
    \boldsymbol{\widetilde{\prodbias}}+\boldsymbol{\widetilde{H}}\left\{( \boldsymbol{\widetilde{U}}\matnot{3}^{T} \widetilde{\boldsymbol{x}}) * \left[\left(\boldsymbol{\widetilde{U}}\matnot{2}^{T}  \widetilde{\boldsymbol{x}}\right) *\left(\boldsymbol{\widetilde{U}}\matnot{1}^{T} \widetilde{\boldsymbol{x}}\right)+\right.\right. 
    \left.\left.\boldsymbol{\widetilde{U}}\matnot{1}^{T} \widetilde{\boldsymbol{x}}\right]+\left(\boldsymbol{\widetilde{U}}\matnot{2}^{T} \widetilde{\boldsymbol{x}}\right) *\left(\boldsymbol{\widetilde{U}}\matnot{1}^{T} \widetilde{\boldsymbol{x}}\right)+\boldsymbol{\widetilde{U}}\matnot{1}^{T} \widetilde{\boldsymbol{x}}\right\},
    \end{split}
    \end{equation}
    which could be expressed as the recursive relationship \eqref{eq:polygan_recursive_gen_third_order_cccp_final}. 

\subsection{Derivations for \biasmodelcccp{}}
\label{ssec:sup_biasmodelcccp}
A polynomial expansion of order $N \in \naturalnum$ with output $\boutvar \in \complexnum^o$ has the form:
\begin{equation}
    \boutvar = \sum_{n=1}^N \;\; \underbrace{\sum_{\sgamma{1} = 2}^{n+1} \;\sum_{\sgamma{2} = \sgamma{1} + 1}^{n+2} \ldots \sum_{\sgamma{N-n} = \sgamma{N-n-1} + 1}^{N}\;}_{(N-n) \text{ sums}} \bigg(\bmcal{\widetilde{W}}^{[n, \sgamma{1}, \sgamma{2}, \ldots, \sgamma{N-n}]} \prod_{j=2}^{n+1} \times_{j} \binvar  \prod_{\tau=n+2}^{N+1} \times_{\tau} \bm{\widetilde{\rho}}\matnot{r_{\tau-n-1}} \bigg) + \bm{\widetilde{h}}\;,
    \label{eq:apollo_poly_general_eq_for_bias_CCP}
\end{equation}
where we add the scaling parameters $\big\{\bm{\widetilde{\rho}}\matnot{n} \in \complexnum^k\big\}_{n=1}^N$.
Then, the tensors $\{\bmcal{\widetilde{W}}^{[n, \sgamma{1}, \sgamma{2}, \ldots, \sgamma{N-n}]} 
\}_{n=1}^{N}
$ and $\big\{\bm{\widetilde{\rho}}\matnot{n}\big\}_{n=1}^N$ are the learnable parameters.
Similarly, to previous cases, the learnable parameters are increasing exponentially, so a standard decomposition will be applied to reduce them. 

Let us rewrite \eqref{eq:apollo_poly_general_eq_for_bias_CCP} for a third-degree polynomial, i.e. $N=3$, to illustrate the decomposition and then we provide the recursive relationship that can be used for an arbitrary degree of expansion. The third degree expansion of \eqref{eq:apollo_poly_general_eq_for_bias_CCP} has the following form:

\begin{equation}
\begin{split}
    \boutvar = \bmcal{\widetilde{W}}^{[1, 2, 3]} \times_{2} \binvar \times_{3} \bm{\widetilde{\rho}}\matnot{2} \times_{4} \bm{\widetilde{\rho}}\matnot{3} + 
    \bmcal{\widetilde{W}}^{[2, 2]} \times_{2} \binvar \times_{3} \binvar \times_{4} \bm{\widetilde{\rho}}\matnot{2} + \\  
    \bmcal{\widetilde{W}}^{[2, 3]} \times_{2} \binvar \times_{3} \binvar \times_{4} \bm{\widetilde{\rho}}\matnot{3} + 
    \bmcal{\widetilde{W}}^{[3]} \times_{2} \binvar \times_{3} \binvar \times_{4} \binvar+\bm{\widetilde{h}}\,.
\end{split}
    \label{eq:apollo_poly_general_eq_for_bias_CCP_third_degree}
\end{equation}
Then, based on \eqref{eq:apollo_poly_general_eq_for_bias_CCP_third_degree}, we can apply a tailored coupled CP decomposition with the following form (in matrix form):
\begin{itemize}
    \item First degree parameters: $\bmcal{\widetilde{W}}^{[1, 2, 3]}_{(1)}:  \bm{\widetilde{H}}(\bm{\widetilde{U}}\matnot{1} \odot \bm{I} \odot \bm{I})^T$.
    \item Second degree parameters: $\bmcal{\widetilde{W}}^{[2, 2]}_{(1)}: \bm{\widetilde{H}}(\bm{\widetilde{U}}\matnot{1} \odot \bm{\widetilde{U}}\matnot{3} \odot \bm{I})^T$.
    \item Second degree parameters: $\bmcal{\widetilde{W}}^{[2, 3]}_{(1)}: \bm{\widetilde{H}}(\bm{\widetilde{U}}\matnot{1} \odot \bm{\widetilde{U}}\matnot{2} \odot \bm{I})^T$.
    \item Third degree parameters: $\bmcal{\widetilde{W}}^{[3]}_{(1)}: \bm{\widetilde{H}}(\bm{\widetilde{U}}\matnot{1} \odot \bm{\widetilde{U}}\matnot{2} \odot \bm{\widetilde{U}}\matnot{3})^T $,
\end{itemize}
where $\bm{I}$ denotes the identity matrix, the parameters $\bm{\widetilde{H}} \in \complexnum^{o\times k}, \bm{\widetilde{U}}\matnot{n} \in  \complexnum^{d\times k}$ for $n=1,2,3$ are learnable.
By plugging in the aforementioned factorization into \eqref{eq:apollo_poly_general_eq_for_bias_CCP_third_degree} and applying Lemma.\ref{lemma:polygan_lemma_hadamard_kr2}, we obtain:
\begin{equation}
    \boutvar =\bm{\widetilde{H}}
    \left(
    (\bm{\widetilde{U}}\matnot{3}^T \binvar + \bm{\widetilde{\rho}}\matnot{3}) *
    (\bm{\widetilde{U}}\matnot{2}^T \binvar + \bm{\widetilde{\rho}}\matnot{2}) * (\bm{\widetilde{U}}\matnot{1}^T \binvar)
    \right)
    + \bm{\widetilde{h}}
    \;.
    \label{eq:apollo_poly_bias_CCP_third_degree_towards_recursive}
\end{equation}
We can generalize the equation above to a general recursive formulation, i.e., \eqref{eq:biasmodelcccp} in the main body.

\subsection{Derivations for \biasmodelmixccp{}}
\label{ssec:sup_for_lccp}
Compared to \biasmodelcccp{}, the scaling parameters in \biasmodelmixccp{} is real-valued instead of complex-valued, i.e., $\big\{\bm{{\rho}}\matnot{n} \in \realnum{}^k\big\}_{n=1}^N$. The third degree expansion has the following form: 
\begin{equation}
\begin{split}
    \boutvar = \bmcal{\widetilde{W}}^{[1, 2, 3]} \times_{2} \binvar \times_{3} \bm{\rho}\matnot{2} \times_{4} \bm{\rho}\matnot{3} + 
    \bmcal{\widetilde{W}}^{[2, 2]} \times_{2} \binvar \times_{3} \binvar \times_{4} \bm{\rho}\matnot{2} + \\  
    \bmcal{\widetilde{W}}^{[2, 3]} \times_{2} \binvar \times_{3} \binvar \times_{4} \bm{\widetilde{\rho}}\matnot{3} + 
    \bmcal{\widetilde{W}}^{[3]} \times_{2} \binvar \times_{3} \binvar \times_{4} \binvar+\bm{\widetilde{h}}\; .
\end{split}
    \label{eq:apollo_poly_general_eq_for_bias_CCPmix_third_degree}
\end{equation}

We apply a tailored coupled CP decomposition with the following form, (note that $\bm{U}\matnot{n}$ is real-valued):
\begin{itemize}
    \item First degree parameters: $\bmcal{\widetilde{W}}^{[1, 2, 3]}_{(1)}:  \bm{\widetilde{H}}(\bm{U}\matnot{1} \odot \bm{I} \odot \bm{I})^T$.
    \item Second degree parameters: $\bmcal{\widetilde{W}}^{[2, 2]}_{(1)}: \bm{\widetilde{H}}(\bm{U}\matnot{1} \odot \bm{U}\matnot{3} \odot \bm{I})^T$.
    \item Second degree parameters: $\bmcal{\widetilde{W}}^{[2, 3]}_{(1)}: \bm{\widetilde{H}}(\bm{U}\matnot{1} \odot \bm{U}\matnot{2} \odot \bm{I})^T$.
    \item Third degree parameters: $\bmcal{\widetilde{W}}^{[3]}_{(1)}: \bm{\widetilde{H}}(\bm{U}\matnot{1} \odot \bm{U}\matnot{2} \odot \bm{U}\matnot{3})^T $,
\end{itemize}
where $\bm{I}$ denotes the identity matrix, the parameters $\bm{\widetilde{H}} \in \complexnum^{o\times k}, \bm{U}\matnot{n} \in  \realnum{}^{d\times k}$ for $n=1,2,3$ are learnable.
By plugging in the aforementioned factorization into \eqref{eq:apollo_poly_general_eq_for_bias_CCPmix_third_degree} and applying Lemma.\ref{lemma:polygan_lemma_hadamard_kr2}, we obtain: 
\begin{equation}
    \boutvar =
    \bm{\widetilde{H}}
    \left(
    (\bm{U}\matnot{3}^T \binvar + \bm{{\rho}}\matnot{3}) *
    (\bm{{U}}\matnot{2}^T \binvar + \bm{{\rho}}\matnot{2}) * (\bm{U}\matnot{1}^T \binvar)
    \right)+
    \bm{\widetilde{h}}
    \;.
\end{equation}
We can generalize the equation above to a general recursive formulation, i.e., \eqref{eq:polygan_recursive_gen_third_order_lccp_final} in the main body.
\section{Experimental details}
This section presents the experimental details, including the description of dataset in Sec
~\ref{sec:dataset}, the quantitative metrics in Sec~\ref{sec:metrics}, and the experimental setup in Sec~\ref{sec:detail_unsupervised_generation},~\ref{sec:detail_condition_generation},~\ref{sec:detail_multimodal}. We train our model on \textbf{a single NVIDIA $2080$ Ti GPU}.
\subsection{Datasets}
\label{sec:dataset}
Firstly, we conduct experiments of unconditional generation on three datasets used in \citet{donahue2018adversarial}: (a) Speech Commands Zero Through Nine (SC09)~\cite{warden2018speech} that consists of spoken digits ``zero'' through ``nine.'' There are $18,620$ audio clips in the training set of SC09 and $2552$ audio clips in the testing set. The duration of each sample is around 1 second. The total length of the training
set is $5.3$ hours. (b) Piano, a music dataset that contains $0.3$ hours' Bach compositions. (c) Drum, a music dataset that contains $0.7$ hours' drum samples in the training set. There are $2350$ audio clips in the training set and $224$ audio clips in the testing set. The duration of each sample is around $1$ second. (d) We conduct the experiment of conditional generation on the same aforementioned SC09 dataset. (e) In addition, we also investigate a large music dataset, called NSynth, that consists $300,000$ musical notes labelled with pitch, instrument, acoustic qualities, and velocity~\cite{nsynth2017}. Each sample lasts for $4$ seconds. We use the same subset and the same test/train split as in \citet{engel2019gansynth}. The labels are the pitches ranging from MIDI $24$ ($\sim$$32$Hz) to MIDI 84 ($\sim$$1000$Hz). 
(f) Note that in the paper of SASHIMI, they use a different splitting for SC09 dataset where there are $31,158$ audio clips in the training set ($8.7$ hours in total). Thus, we train our model again in this training set to fairly compare with the baselines, as reported in Table~\ref{tab:poly_exper_audio_IS_diffusion}.

\subsection{Evaluation metrics in audio generation}
\label{sec:metrics}
\begin{figure}[!h]
    \centering
    \includegraphics[width=0.6\linewidth]{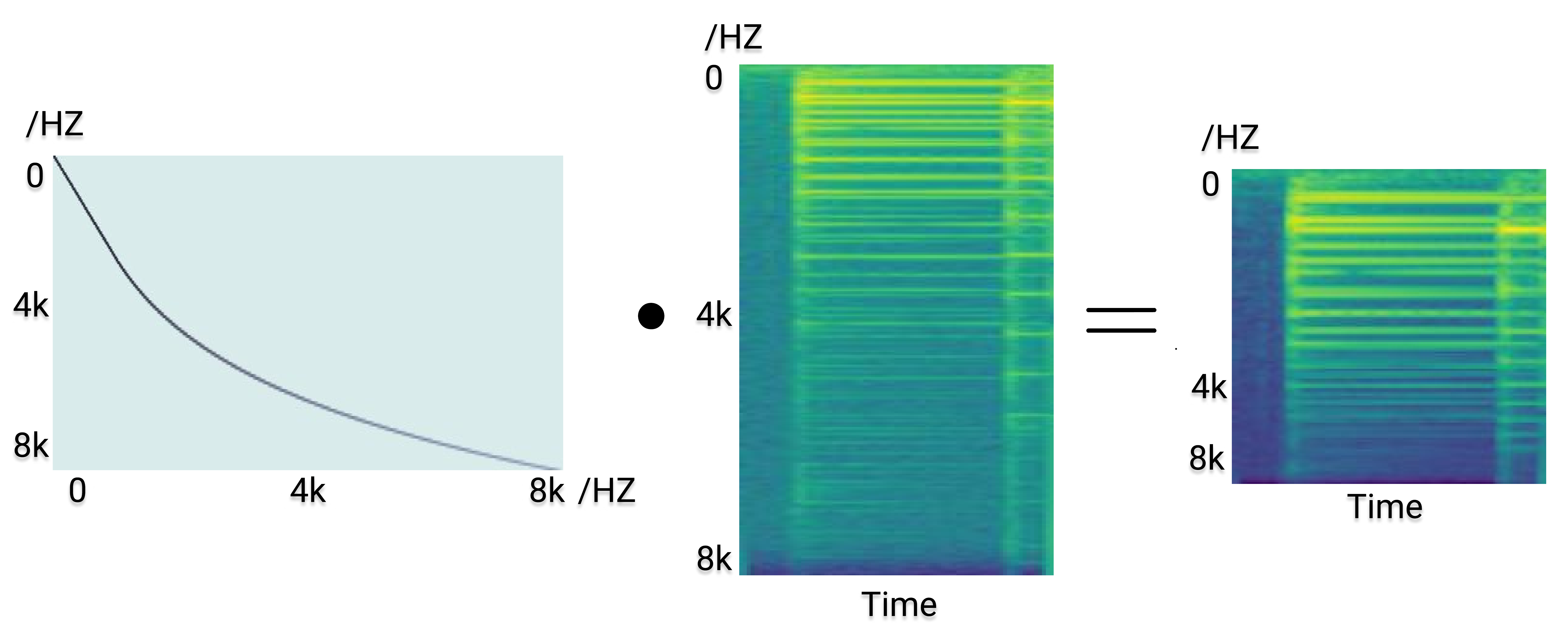}\vspace{-2mm}
   \caption{Before calculating JSD, NDB, IS, and FID, 
    we need to project the magnitude of STFT  (\textbf{Middle}) to human perceptual frequency scales (\textbf{Right})  through mel-scale projection matrix  (\textbf{Left}).
The spectrum in the figure are plotted in log-scale only for visualization purpose.
    }
\label{fig:mel}
\end{figure}

\begin{itemize}
    \item 
    \textbf{Inception Score (IS).} Firstly introduced by ~\citet{NIPS2016_8a3363ab}, IS is used to measure the quality and the diversity of the image generated by GAN. The fake samples are fed into a pretrained Inception Network V3~\cite{Szegedy_2015_CVPR} to get the conditional label distribution. The IS is calculated by averaging the KL divergence between the conditional label distribution and its marginal distribution. 
    The audio samples in SC09 dataset are converted to log spectrum representation with $16$k sample rate, $8$ ms stride, and $64$ ms windows size. The frequency bin is further mapped into mel-scale ranging from $40$ HZ to $7800$ HZ.
    We use the same evaluation protocol ($50$k samples) and the same pre-trained classifier from the paper of WaveGAN~\citep{donahue2018adversarial}to calculate the IS for SC09 dataset.
    
    \item \textbf{Fréchet Inception Distance (FID)}.
    FID ~\cite{NIPS2017_8a1d6947}
    characterizes the Fréchet distance 
    of the intermediate layer's feature from the pretrained Inception Network V3 between the real image and the generated image.
    For SC09 dataset, 
   the evaluation protocol and the pre-trained classifier are the same as that in IS.
     For NSynth dataset, the audio is converted to log spectrum representation with $16$k sample rate, $16$ ms stride, and $128$ ms windows size. The frequency bin is further mapped into mel-scale ranging from
    $0$ Hz to $8000$ Hz. we train a CNN-based pitch classifier to calculate the score.
     \item \textbf{Number of Statistically-Different Bins (NDB) and Jensen-Shannon Divergence
(JSD)}.
     NDB and JSD are proposed by ~\citet{10.5555/3327345.3327486} to measure the diversity between the generated samples and the real samples.
    The audio is converted to STFT and then mel-spectrogram, as illustrated in Fig~\ref{fig:mel}.
    For SC09 (Piano, Drum, NSynth) dataset,
   the mel-scale is ranging from
   40 Hz to 7800 Hz (0 HZ to 8000 HZ). The mel-spectrograms of the real samples are clustered by K-means into 50 clusters.
    \item 
   Unless otherwise mentioned, all evaluation metrics in this paper are based on the aforementioned details. However, due to the popularity of audio synthesis, a variety of techniques have been proposed for evaluation. Specifically, in the comparison with non-adversarial methods, as shown in Table~\ref{tab:poly_exper_audio_IS_diffusion}, the evaluation protocol follows the paper of SASHIMI \citep{goel2022s}, which was released within the last few months. We use the same  pretrained-model, number of samples (2048), and metrics (IS, Modified Inception Score~\citep{gurumurthy2017deligan}, FID, AM Score~\citep{zhou2018activation}). \textbf{Modified Inception Score (MIS)}: MIS considers both intra-class and inter-class sample diversity.
    \textbf{AM Score}:
     The difference from IS is that AM Score considers the marginal target distribution of the training set.
    \end{itemize}

\subsection{Experimental setup in unconditional audio generation}
\label{sec:detail_unsupervised_generation}
Each audio clip in all datasets is first down-sampled with $16$kHz and padded or clipped to the fixed length of $16384$.
For the SC09 dataset and Drum dataset, we apply STFT with $8$ ms stride, $16$ ms windows size, and Hann window, which results in the resolution $129\times 128$ (Frequency $\times$ Time). 
We use a resolution of $128\times 128$ by truncating the Nyquist bin, i.e., bottom row. The APOLLO is implemented by the product of \biasmodelrncp{} and \biasmodelcncp{}. Since the sounds in Piano dataset are not sparse in time domain, we choose $8$ ms stride and $32$ ms windows. Similarly, we drop the Nyquist bin and get the final solution $256\times 128$ (Frequency $\times$ Time).  We choose the product of \biasmodelrccp{} and
\biasmodelcncp{}. Note that the matrices used to multiply by the noise vector are implemented by dense layer while the matrices used to multiply by the time-frequency representation is implemented by convolution layer. The Hadamard product is implemented by element-wise multiplication. For the hyperparameters, the factor of the gradient penalty is $10$. In each step, we apply $5$ updates
of the discriminator and 1 update of the generator with learning rate
$10^\text{-4}$.
The hyperparameters of ADAM optimizer~\cite{kingma2014adam} are $\beta_{\text{1}}$ = $0.5$ and $\beta_{\text{2}}$ = $0.9$. The base channel of the CNN is $64$ in both generator and discriminator ($16$ for the 'Small' model). The batch size is $8$.
The model converges within $380$k steps ($2$ days) in SC09 dataset and $100$k steps on Piano dataset and Drum dataset. 
\subsection{Experimental setup in conditional audio generation}
\label{sec:detail_condition_generation}
Fig~\ref{fig:prodpoly_model_conditional} is a schematic illustration of applying \eqref{eq:complex_poly_model1_rec} on the conditional generator. Motivated by \citet{miyato2018cgans}, we use a projection discriminator to incorporate the label information.
For SC09 dataset, other settings are the same as in unconditional generation. Our model converges within $3$ days ($450$k steps). The log spectrum of the generated samples from 0 to 9 are shown in Fig~\ref{fig:prodpoly_model_conditional_sampels}. For NSynth dataset, we choose $16$kHz sample rate and pad (or clip) each clip to the fixed length of $64000$. Then we apply STFT with $16$ ms stride, $128$ ms windows size, and $75$\% overlap, which results in the resolution $1025\times 128$ (Frequency $\times$ Time).
After truncating the Nyquist bin, the final resolution is $1024\times 128$. 
The APOLLO is implemented by the product of \biasmodelrncp{} and \biasmodelcncp{}.
The base channel of CNN in our model is $16$ in this experiment. Other settings are the same as in conditional generation on SC09. Our model converges within $15$ hours ($60$k Steps).
\subsection{Experimental setup in multimodal generation}
\label{sec:detail_multimodal}
 The architecture of the generator is presented in Fig~\ref{fig:image2audio} in the main body, 
 Firstly, we use two low-degree \compoly{}s for the random noise and the image. 
 Specifically, we use \biasmodelrncp{} for the real-valued image and \biasmodelcncp{} for the complex-valued noise. 
 Then, the conditional \compoly{}, i.e., \biasmodelcncp{}, will receive the output of these two low-degree \compoly{}s and generate the STFT of the audio. As for the discriminator,
 we resize the image with nearest-neighbor interpolation to the resolution $128\times128$  and concatenate it with the STFT of the audio at the input layer. The base channel of CNN is $16$ in this experiment.
 Other settings are the same as in unconditional generation on SC09.
 Our model converges within one day ($200$k steps).

\section{Ablation experiments}
\label{sec:ablation}

\subsection{Model variants}
We conduct an ablation study on different schemes proposed in this paper. The quantitative results
are presented in Table~\ref{tab:compelx_polygan_audio_diffmodel}, where all models are implemented using products of polynomials, i.e., several polynomials stacked sequentially. '\biasmodelrncp{} + \biasmodelcncp{}' obtains the best result in IS, NDB, and JSD while '\biasmodelrccp{} + \biasmodelrncp{}' achieves the lowest FID. In practice, the model with full decomposition on complex-valued coefficients ('\biasmodelcncp{}+ \biasmodelcncp{}') requires much more parameters, while the improvement is not significant.
\begin{table*}[!h]
    \centering
    \caption{Ablation experiment on different proposed schemes of \compoly{}.
     }
     \begin{tabular}{|c | c|c|c|c|c|} 
     \hline
     \multicolumn{6}{|c|}{Unconditional audio generation on SC09 dataset}\\ 
     \hline
     Model & IS ($\uparrow$) & FID ($\downarrow$)  & NDB ($\downarrow$)  & JSD ($\downarrow$) & \# par\\
    \hline
     Real data& $8.01 \pm 0.24$  &$0.50$ &$0.00\pm0.00$ & $0.011$&$\_$ \\
\hline

\biasmodelmixccp{}+\biasmodelmixncp{}
& $6.74\pm0.06$ & $9.34$&$3.80\pm0.74$ &$0.039$ & $46.0$\\

\biasmodelrccp{}+\biasmodelmixncp{}
& $6.98\pm0.04$ & $12.00$&$4.00\pm0.00$ &$0.040$ & $45.9$\\

\biasmodelrccp{}+\biasmodelrncp{}
&$6.79\pm 0.03$ &$\boldsymbol{8.03}$ &$3.80\pm1.32$ &$0.041$& 45.9\\

\biasmodelrncp{}+\biasmodelrncp{} 
& $6.71\pm0.03$ 
& $9.44$  
& $4.00\pm0.00$ 
&$0.038$ & 45.9
\\
\biasmodelrncp{}+\biasmodelcncp{} &$\boldsymbol{7.25\pm0.05}$ 
&$8.15$ 
&$\boldsymbol{3.20\pm1.16}$ &$\boldsymbol{0.029}$& 64.1\\

\biasmodelcncp{}+\biasmodelcncp{}
&$6.84\pm 0.02$ &$8.87$ &$4.20\pm0.40$ &$0.041$& 68.1\\

\hline
\end{tabular}
\label{tab:compelx_polygan_audio_diffmodel}
\end{table*}

\subsection{The field ($\complexnum$ or $\realnum$) of the generator and the discriminator}
Below, we conduct experiment with different fields of the generator and discriminator.
We combine our \compoly{} (\biasmodelrncp{}+\biasmodelcncp{}), with real-valued discriminator and complex-valued discriminator respectively. Additionally, \pinet{} is tested as a real-valued generator that concatenates the real part and the imaginary part of the STFT in two channels. The result in Table~\ref{tab:compelx_polygan_audio_difffield} illustrates that our model~\compoly{} with real-valued discriminator yields best
IS and FID.
 \begin{table}[h]
    \centering
    \caption{Ablation experiment on the field ($\complexnum$ or $\realnum$) of generator and discriminator (D). The results demonstrate that combining our complex-valued generator~\compoly{} with real-valued discriminator yields the best performance in practice.
    }
     \begin{tabular}{|c | c|c|} 
     \hline
     \multicolumn{3}{|c|}{Unconditional audio generation on SC09 dataset}\\ 
     \hline
     Model & IS ($\uparrow$) & FID ($\downarrow$) \\
    \hline
     Real data& $8.01 \pm 0.24$  &$0.50$\\
     \hline
    \pinet{}-real D& ${6.59\pm0.03}$ & $13.01$\\      \hline
    \compoly{}-real D&$\boldsymbol{7.25\pm0.05}$ &$\boldsymbol{8.15}$ \\ 
    \compoly{}-complex D& ${6.74\pm0.05}$ & $11.76$\\ \hline
     \end{tabular}
     \label{tab:compelx_polygan_audio_difffield}
\end{table}

\subsection{Removing activation functions on audio generation}
\begin{figure}[htb]
    \centering
    \includegraphics[width=0.5\linewidth]{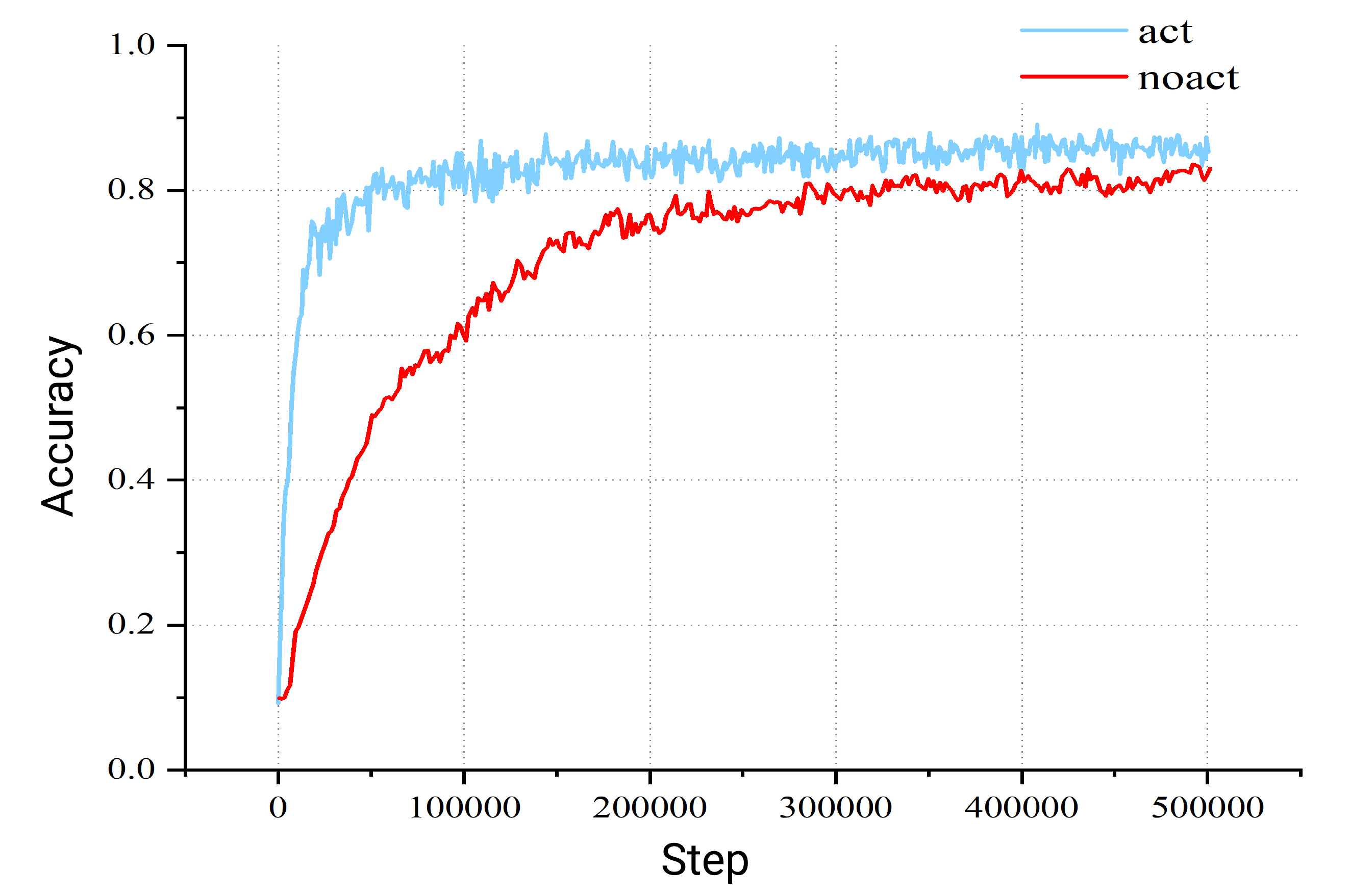} \vspace{-3mm}
\caption{Ablation experiment on removing the activation functions between the layers in the task of conditional audio generation. The figure plots the accuracy of the classes of the generated samples given the label inputs during the training process. The model without activation functions converges more slowly and obtains a lower accuracy.}
\label{fig:acc}
\end{figure}

In practice, we add ReLU activations function after all Hadamard products in the implementation, which converts the network into a piece-wise polynomial expansion. In this ablation experiment, we maintain the same setting as in previous experiment except for the generator. Particularly,
the generator only contains a hyperbolic
tangent in the output space for normalization as typically done in other GAN generators.
There is no activation function between the layers. The result in Table~\ref{tab:poly_exper_audio_sc09_con_noact} shows that both metrics deteriorate without using
non-linear activation function.
We use the pretrained classifier to calculate the accuracy of the generated samples given the label inputs. Fig~\ref{fig:acc} shows that the model converges slower and obtains lower accuracy notably after removing non-linear activation functions.

\begin{table}[htb]
    \centering
    \caption{Ablation experiment  on removing activation function between the layers in the task of conditional generation on SC09. The result shows that our model can be trained without using activation functions (denoted as 'noact'). However, inserting such non-linear operator (denoted as 'act') in each degree significantly boosts the performance in practice.
    }
     \begin{tabular}{|c | c|c|} 
     \hline
     \multicolumn{3}{|c|}{Conditional audio generation on SC09 dataset}\\ 
     \hline
     Model & IS ($\uparrow$)  & FID ($\downarrow$) \\
    \hline
     \compoly{}-act& $\boldsymbol{7.73\pm0.04} $& $ \boldsymbol{ 6.31}$\\
     \compoly{}-noact& $6.34\pm0.03 $& $ 15.90$\\
  
      \hline
     \end{tabular}
     \label{tab:poly_exper_audio_sc09_con_noact}
\end{table}

\subsection{Removing activation functions on speech recognition}
We have demonstrated the feasibility of removing the activation functions of our model in generative task, in this section, we conduct a similar experiment on audio recognition. That is, we remove the activation functions between the layers. The result in Table~\ref{tab:classification_cqt_withoutact} shows that our model can obtain similar accuracy even though we remove all the activation functions.
\begin{table}[!h]
 \caption{Ablation experiment on removing activation function between the layers in the task of speech recognition, the result shows that our model can be trained without using activation functions (denoted as 'noact'). 
 Inserting the non linear ReLU activations function (denoted as 'act') only slightly improves the accuracy.}
\centering
     \begin{tabular}{|c | c |}
         \hline
         \multicolumn{2}{|c|}{ Classification on Speech Commands Dataset}\\ 
         \hline
         Model   & Accuracy\\
        \hline
              \compoly{}-act  & $\boldsymbol{0.923}$\\
              \compoly{}-noact  & $0.920$\\
         \hline
     \end{tabular}
 \label{tab:classification_cqt_withoutact}
\end{table}
\subsection{Qualitative evaluation through interpolation}
\label{ssec:complex_poly_interpolation}

In this section, we conduct qualitative evaluations of the model through interpolation. We first consider the model of Sec~\ref{sssec:complex_poly_method_two_variable}, i.e., the conditional generation experiment on SC09 dataset. As a reminder, the generator accepts the latent codes, and the class-conditional label (digit). 

\begin{figure*}[!h]
\centering
    \includegraphics[width=0.9\linewidth]{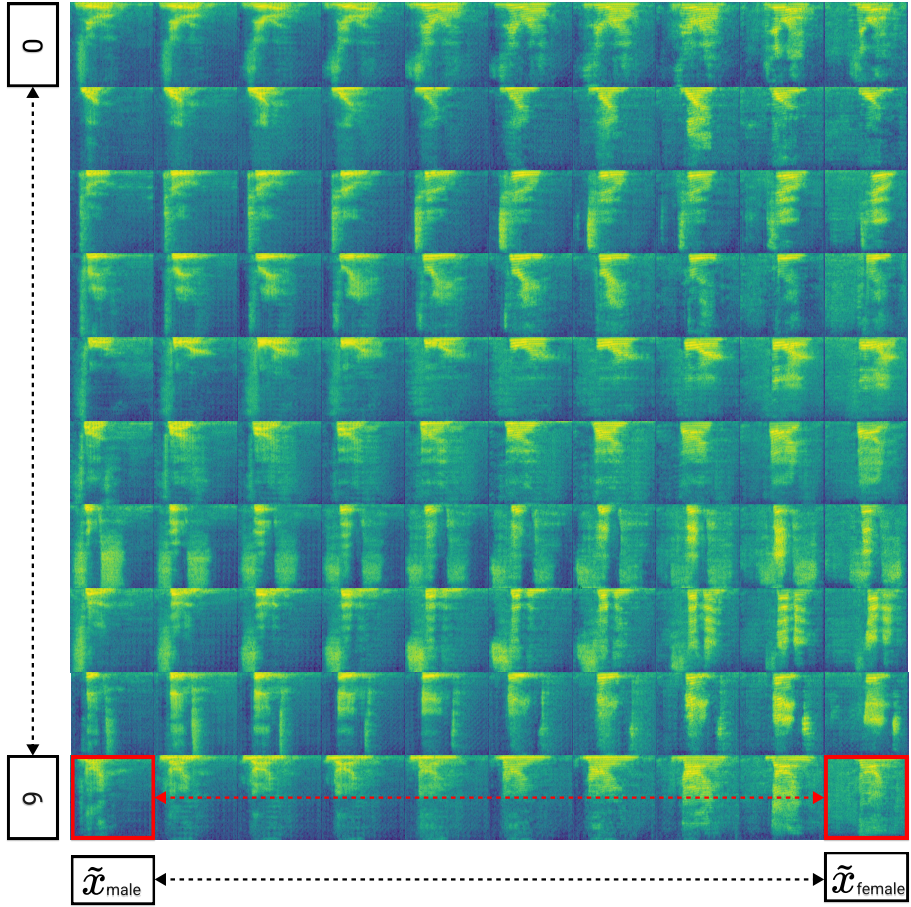}
\caption{The figure shows the log spectrum of the samples generated by interpolation in the class-conditional model. The full description of the interpolation is on Sec~\ref{ssec:complex_poly_interpolation}. As a remark, the male (female) attributed voice was annotated by a human expert on synthesized audio samples on digit '9' with random noise $\widetilde{x}_\text{male}$ ($\widetilde{x}_\text{female}$), i.e., the one demonstrated with red. The rest digits are synthesized by interpolating between the two latent codes along with the digit-based labels, i.e., 0-9 digits.}
\label{fig:complex_poly_model_interpolation_sampels}
\end{figure*}

We fix the class-conditional label, e.g., to digit '9'. We sample two latent codes and we manually annotate the corresponding synthesized audio samples with respect to the gender. We select two latent codes that correspond to a female-attributed voice and a male-attributed voice\footnote{The attribution of gender on the voice is a significant topic that we do not focus on this work; our experiment relies on the annotation of an expert solely for demonstration purpose.} . We then apply a linear interpolation between the two latent codes, which results in ten discrete latent codes. We then use these latent codes and vary the class-conditional label from '0' to '9'. The resulting log spectrum is visualized in Fig~\ref{fig:complex_poly_model_interpolation_sampels}, where we observe a smooth transition in the log spectrum space. Importantly, the audio samples demonstrate how the voice varies from the beginning of each row (male-attributed voice) to the end of each row (female-attributed voice). 

\begin{figure*}[!h]
    \centering
    \includegraphics[width=0.9\linewidth]{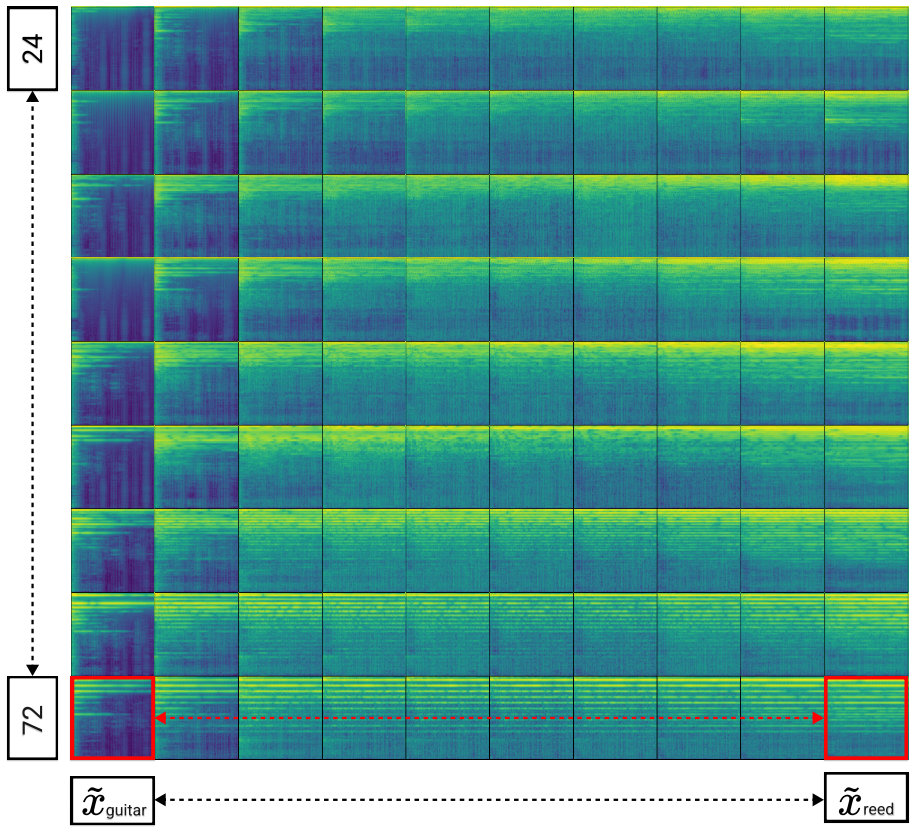}
\caption{The figure depicts the log spectrum of the samples generated by interpolation in the class-conditional model trained on NSynth dataset. For each small image, the horizontal (vertical) axis is along the time
(frequency), the frequency increases with interval scale from top to bottom. The full description of the interpolation is on Sec~\ref{ssec:complex_poly_interpolation}. As a remark, the guitar (reed) attributed musical instrument sounds was annotated by a human expert on synthesized audio samples on pitch '$72$' with random noise $\widetilde{x}_\text{guitar}$ ($\widetilde{x}_\text{reed}$), i.e., the one demonstrated with red. The rest digits are synthesized by interpolating between the two latent codes along with the pitch-based labels from $24$ to $72$ with interval scale.}
\label{fig:complex_poly_model_interpolation_sampels_nsynth}
\end{figure*}

We also conduct the interpolation for the model trained on NSynth datatset. We select two latent codes that correspond to a guitar-attributed audio and a reed-attributed audio. Then we apply the linear interpolation between these two latent codes and vary the pitch from $72$ to $24$. The resulting log spectrum is visualized in
Fig~\ref{fig:complex_poly_model_interpolation_sampels_nsynth}, where we observe a smooth transition in the log
spectrum space. Notably, as the pitch decreases, the low-frequnecy component in the log spectrum becomes more prominent.
\subsection{Similarity of representations between different degrees}
\label{ssec:complex_poly_similarity_representations}
To further investigate the properties of the \compoly, we explore the correlations captured by representations across layers. To that end, we rely on Canonical Correlation Analysis (CCA)~\cite{NIPS2017_7188,CCA} to analyze the representations between the different degrees (i.e., layers). 
Concretely, CCA aims to maximize the correlation between two different sets of variates.
In our case, we treat the value after the Hadamard product as the representation of each degree.
Since the spatial dimensions of the output of different degrees are not the same, we interpolate the smaller one to match the dimensions. The open-source implementation of singular value CCA~\cite{NIPS2017_7188,CCA} is utilized.

\begin{figure}[htb]
    \centering
    \includegraphics[width=0.7\linewidth]{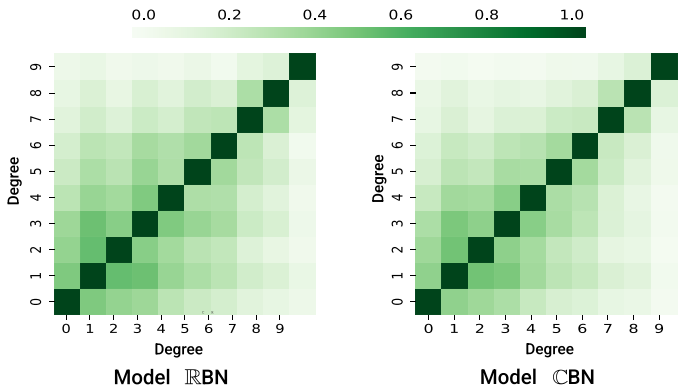} \vspace{-2mm}
\caption{The similarities between the outputs. The color-scale on the top ranges from $[0, 1]$ and dictates the correlation.}
\label{fig:CCA}
\end{figure}

As a case study, we use the models of \biasmodelrncp{} and \biasmodelmixncp{} that have similar decompositions and differ in the parameters on the last layer. 
The results are depicted in Fig~\ref{fig:CCA}, where we observe that the behavior in different layers of \biasmodelrncp{} and \biasmodelmixncp{} is similar. Namely, the representations of different degrees seem to have a localized behavior with the correlations between early-layer representations and last-layer representations to be low. 

\section{Additional experiments}
\label{sec:complex_polynomial_appendix_experiments}

\subsection{Inference speed}
\label{sec:appendixinferencespeed}
The result is presented in Table~\ref{tab:inferencespeed}.
WaveGAN has the highest inference speed since it synthesizes the audio directly in the time domain.
Notice that the complex operations, e.g., complex multiplication, result in an augmented inference time for the proposed \compoly. However, the TiFGAN that scored as the top audio model, after our model and the generic \pinet, in the human study is more computationally demanding than the proposed model. A future step consists in further improving the small model and making the complex operations more efficient.
As for non-adversarial models, we add the comparison with the SOTA model DiffWave and omit SASHIMI, which is implemented by integrating into DiffWave, as mentioned in \citet{goel2022s}. We can see DiffWave is the slowest model due to the reverse process.

\subsection{Human evaluation}
 \label{sec:humanevaluation}
    We collect $30$ generated samples for each model and $30$ real samples. Then we divide these samples into $3$ groups and concatenate $10$ samples generated by the same model as $1$ audio clip which lasts $10$ seconds. We invite $25$ volunteers whose medium of study are English for the comparison with non-adversarial methods and adversarial methods, respectively. The volunteers are asked to assign an ordinal-scale score ($5$: excellent, $4$: good, $3$: fair, $2$: poor, $1$: bad) to each audio clip based on the sound quality and perceptibility.
    Finally, we calculate the mean for each model, which is as known as mean opinion score (MOS)~\citep{Ribeiro2011CROWDMOSAA}.
\begin{table}[!th]
\setlength{\tabcolsep}{4pt}
    \centering
    \caption{
    This table presents the inference speed for those models unconditionally trained on SC09.
    '\#spb' abbreviates the seconds per batch (batch size = 128) during inference on a single NVIDIA 2080 Ti GPU.
'\# par’ abbreviates the number of parameters.
} 
\scalebox{0.9}{
    \begin{tabular}{|c | c|c|c|c|c|} 
         \hline
        \multirow{2}{*}{Model} & Frequency   & Additional phase  &GAN& {\#spb}        & {\# par}\\ 
         & domain & recovery technique& based&($\downarrow$)&($\downarrow$)\\
        \hline  
        WaveGAN  &\xmark&\xmark&\colorcheck&$\boldsymbol{0.039}$&$36.5$ \\\hline
        SpecGAN &\colorcheck&\colorcheck&\colorcheck&$0.069$&$36.5$\\ \hline
        TiFGAN &\colorcheck&\colorcheck&\colorcheck&$4.926$&$43.4$\\   \hline
        DiffWave &\xmark&\xmark&\xmark&$>480$&$43.4$\\   \hline
        \pinet{} &\colorcheck&\xmark&\colorcheck &$0.067$&$45.9$ \\ \hline
        \multicolumn{5}{|c|}{\hspace{1.3cm}\compoly}\\ 
        \biasmodelrccp{}+\biasmodelmixncp{}, Small &\colorcheck&\xmark&\colorcheck &$0.072$& $\boldsymbol{3.5}$\\
        \biasmodelrccp{}+\biasmodelmixncp{}&\colorcheck&\xmark&\colorcheck & $0.315$& $45.9$\\
        \biasmodelrncp{}+\biasmodelcncp{}, Small&\colorcheck&\xmark&\colorcheck & $0.091$& $4.6$\\
        \biasmodelrncp{}+\biasmodelcncp{} &\colorcheck&\xmark&\colorcheck  &0.374& $64.1$\\
        \hline
    \end{tabular}
}
\label{tab:inferencespeed}
\end{table}
\begin{figure}[!htb]
    \centering
     \begin{subfigure}[b]{0.49\textwidth}
         \centering \includegraphics[width=1\textwidth]{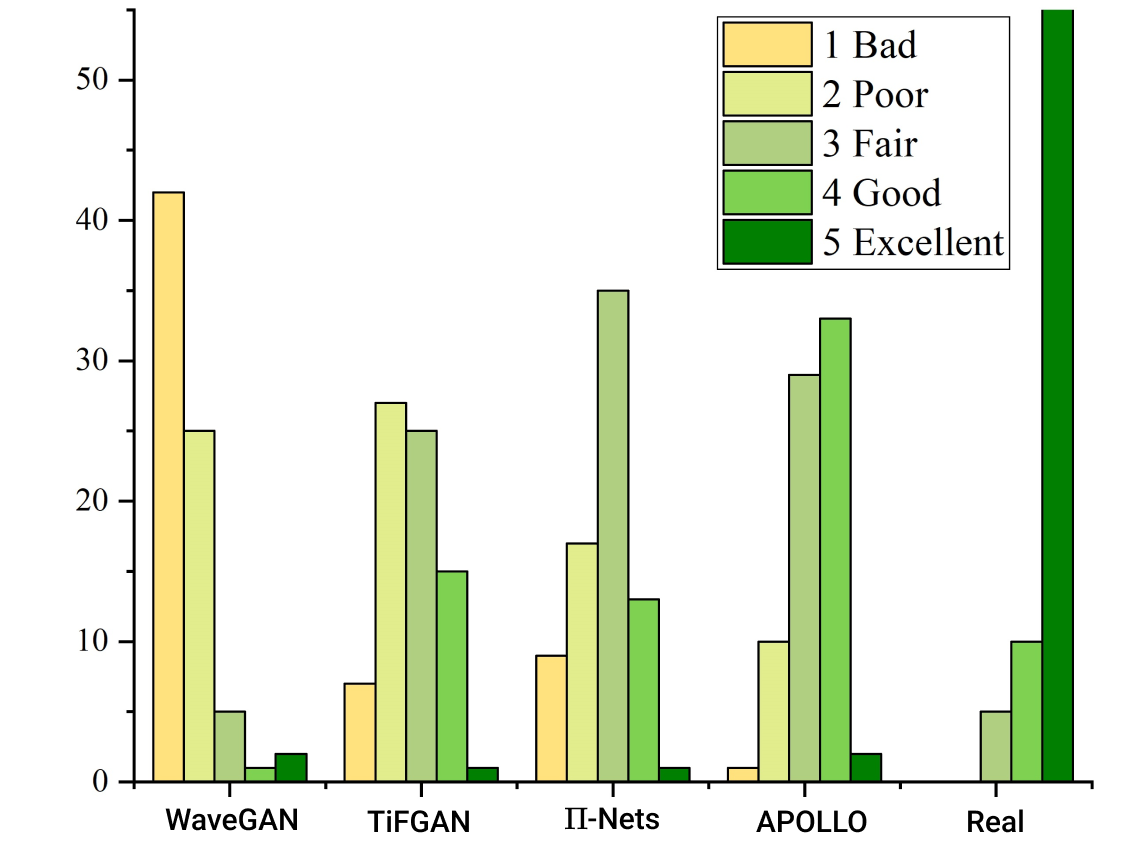}
         \caption{
         This histogram corresponds the experiment in Sec~\ref{ssec:Unsupervised_audio_generation} and shows human evaluation result with our method, adversarial methods, and real data.
         From left to right,
         the MOS for all models and real data are $1.61, 2.68, 2.73, 3.33,4.73$, respectively. 
}
     \end{subfigure}
     \hfill
     \begin{subfigure}[b]{0.49\textwidth}
         \centering
     \includegraphics[width=1\textwidth]{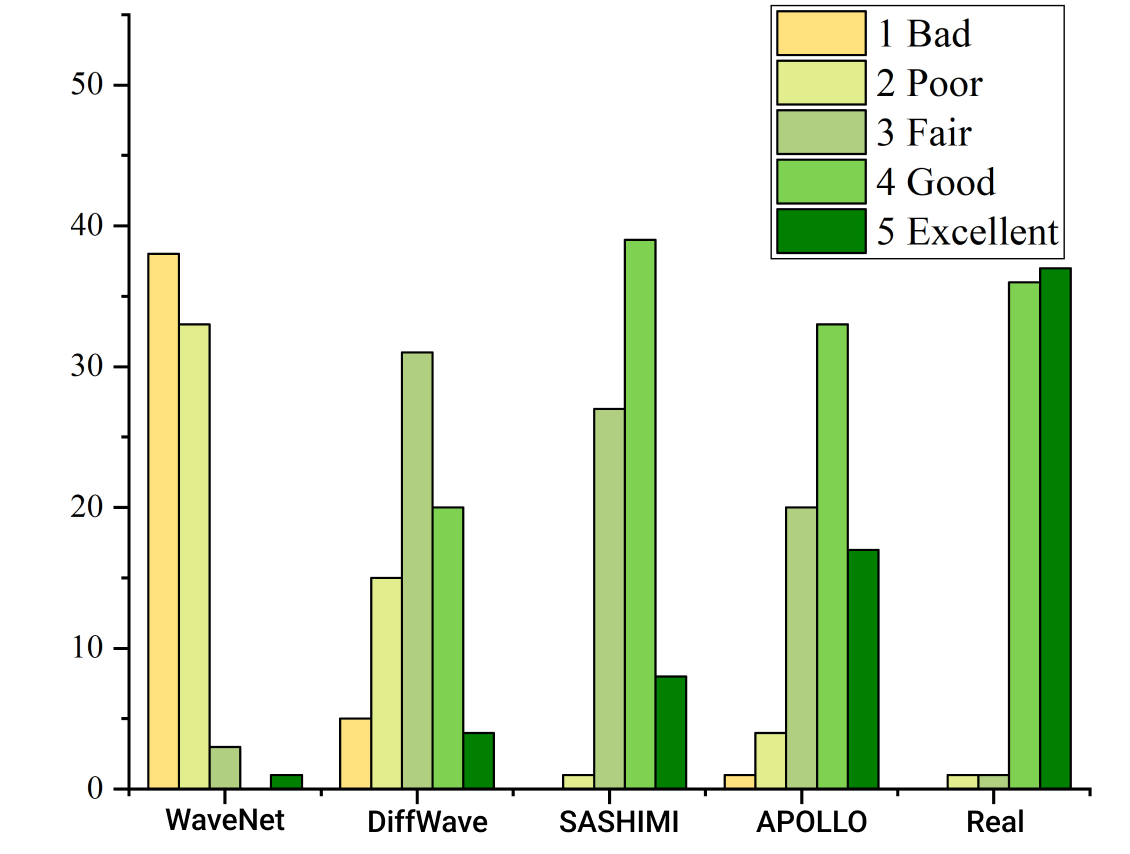}
         \caption{
         This histogram corresponds the experiment in Sec~\ref{ssec:Unsupervised_audio_generation_nonadversarial} and shows human evaluation results with our method, non-adversarial methods, and real data.
         From left to right
         , the MOS for all models and real data are $1.57, 3.04,3.72, 3.81,4.45$, respectively. 
         }
     \end{subfigure}
\caption{Human evaluation on unconditional audio generation on SC09 dataset.
The results validate the enhanced performance of \compoly{} already indicated by the standard quantitative metrics.}
\label{fig:human}
\end{figure}

\subsection{Beyond audio generation}
\label{sec:appendix_beyond_generation}
This section includes additional small-scale experiments on speech recognition and speech enhancement, aiming to show that our networks architecture with the corresponding complex-valued audio representation can also be adapted to other audio-related tasks and achieve considerable performance, which allows future researchers to incorporate our architecture and representations. 

\subsubsection{Audio classification}
\label{sec:detail_audio_classification}
We assess the performance in speech recognition on the Speech Commands Dataset~\cite{warden2018speech}, which consists of 35 different classes.
We choose \resnet{}~\cite{7780459}, its variants          Complex~\resnet~\cite{trabelsi2018deep}, and real-valued polynomial networks(~\pinet)~\cite{chrysos2020p}
         as the baseline. 
 CQT is chosen as the audio representation, where is primitively log-scale along the frequency axis. The inputs of all complex-valued networks are the CQT representation while the inputs of all real-valued networks are the magnitude of the CQT representation. The CQT spans $6$ octaves with $32$ bins per octave. 
The sample rate is $16000$ and the hop-length is $512$. All models are optimized via SGD with momentum
$0.9$, weight decay $5\times10^{-4}$. A batch size of $128$ is
used. The initial learning rate is set to $0.1$ and decreased by a factor of $10$ at $40, 60, 80$, and $100$ epochs. Each model is trained for $120$ epochs. The channels in each model are $[64, 128, 256, 256]$. To scrutinize whether our model could still outperform the baselines, we decrease the channels of our model to $[64, 96, 128, 256]$, which is referred to 'Small' model. The results are summarized in Table \ref{tab:classification_cqt}. The \compoly{} has parameters comparable to the strongest baseline, but outperforms all the baselines by a considerable margin. The small \compoly{} still outperforms the baseline while reducing the parameters by more than $38\%$.
\begin{table}[htb]
 \caption{Speech classification with our model and different \resnet{} variants on Speech Commands Dataset.
 Our best model improves the accuracy by $0.6\%$ while the small model still outperforms the baselines with much fewer parameters.}
\centering
     \begin{tabular}{|c | c | c|}
         \hline
         \multicolumn{3}{|c|}{ Classification on Speech Commands Dataset with CQT}\\ 
         \hline
         Model   & Accuracy & \#par (M)\\
        \hline
         \resnet18 
       & $0.904$   & $11.8$ \\
         \hline
         Complex \resnet18 
        & $0.917$  &  $10.4$ \\
         \hline
          \pinet~\resnet18 
          & $0.912$ &  $6.0$ \\
          \hline
          \compoly{}, Small & ${0.919}$   & $\boldsymbol{3.7}$\\
              \compoly{}  & $\boldsymbol{0.923}$ & $6.3$\\
         \hline
     \end{tabular}
 \label{tab:classification_cqt}
\end{table}
\subsubsection{Speech enhancement}
Next, we conduct a small-scale speech enhancement on VoiceBank-DEMAND dataset~\citep{thiemann2013diverse,veaux2013voice}.
Similar to the previous recognition experiment, we choose baselines that have similar architecture but perform in different fields ($\realnum{}$ or $\complexnum{}$), including (1) Wave-U-Net~\citep{macartney2018improved}, a real-valued U-Net architecture for the waveform representation. (2) DCUnet~\citep{choi2018phaseaware}, a complex-valued U-Net architecture for STFT representation in time-frequency domain.
For our model, we convert the decoder of DCUnet to APOLLO.
The audios in the training set and testing set are firstly downsampled to $16$kHz.
We apply STFT with $64$ms window size and $16$ms hop length.  
The speech quality is evaluated via perceptual evaluation of speech quality (PESQ)~\citep{rix2001perceptual} and MOS predictor of signal distortion (CSIG)~\citep{hu2007evaluation}. Result in Table~\ref{tab:speechenhancement} shows that APOLLO outperforms both baselines, which can be attributed to the expressivity of our complex-valued polynomial networks and corresponding architecture design.
\begin{table}[htb]
 \caption{Speech classification with our model and different U-net variants. The result demonstrates the increasing performance when using our APOLLO.}
\centering
\begin{tabular}{|c|c|c|}
 \hline
 \multicolumn{3}{|c|}{ Speech enhancement on  VoiceBank-DEMAND dataset}\\ 
 \hline
 Model &PESQ($\uparrow$) &CSIG ($\uparrow$)\\
\hline  
Wave-U-Net&2.40&3.52\\
\hline
DCUnet&3.24&4.34\\
  \hline
 \compoly{}&$\boldsymbol{3.32}$&$\boldsymbol{4.53}$\\
  \hline
 \end{tabular}
\label{tab:speechenhancement}
\end{table}

\section{Societal impact}
\label{sec:complex_polynomial_societal_impact}
Our work uses polynomial expansions in the complex fields for audio processing tasks. Among the tasks, we utilize our method for synthesizing audio samples. Audio generation is a significant task with tremendous applications, e.g., human-computer interface. As such, audio generation methods could be used for creating misleading content and we believe that further research is required to ensure the capabilities are used for creating a positive societal impact.

The Generative Adversarial Nets (GANs)~\citep{NIPS2014_5ca3e9b1} that we use for our experiments on audio generation have a dedicated module, i.e., the discriminator, for detecting the real from the fake content. Unfortunately, this is not sufficient for detecting fake content. To that end, we encourage the community to further investigate techniques for discriminating the real from the synthesized audio. Even though our work relies on public benchmarks that are widely used for audio processing, we hope that further research is conducted on how to avoid the negative societal impact from synthesized content.

\end{document}